\documentclass[10pt,conference]{IEEEtran}
\usepackage{booktabs} %
\DeclareMathAlphabet{\mathcal}{OMS}{cmsy}{m}{n}
\usepackage{setspace}
\usepackage[italic]{mathastext}
\usepackage{array}
\usepackage{soul}
\usepackage{titlesec}
\usepackage{fancyhdr}
\usepackage[normalem]{ulem}
\usepackage{datetime} 
\usepackage{authblk}
\makeatletter
\renewcommand\AB@affilsepx{ \protect\Affilfont}
\makeatother
\usepackage{multirow}
\usepackage{multicol}
\usepackage{listings}
\usepackage{color}
\usepackage[font={small,bf}]{caption}
\usepackage{float}
\usepackage[nomessages]{fp}
\usepackage{dblfloatfix}    %
\usepackage{bm}
\usepackage[noadjust]{cite}
\usepackage{tikz}
\usepackage{xcolor}
\usepackage[font={small,bf}]{subcaption}
\usepackage[linesnumbered,ruled]{algorithm2e}
\usepackage{makecell}

\usepackage{todonotes} %
\usepackage{marginnote} 

\usepackage{listings}
\usepackage{amssymb}
\usepackage{amsmath}

\usepackage[]{microtype}

\usepackage[utf8]{inputenc}
\usepackage[T1]{fontenc}

\usepackage{mathptmx} %

\def\BibTeX{{\rm B\kern-.05em{\sc i\kern-.025em b}\kern-.08em
    T\kern-.1667em\lower.7ex\hbox{E}\kern-.125emX}}

\newcommand{\myName}{NEON\xspace}
\newcommand{\guestName}{workload\xspace}
\newcommand{\myTitle}{Enabling Efficient Support for\\Nonlinear Operations in Resistive RAM-based\\Neural Network Accelerators\xspace}

\newcommand\redout{\bgroup\markoverwith{\textcolor{red}{\rule[0.5ex]{2pt}{1pt}}}\ULon}

\definecolor{dblue}{rgb}{0.00, 0.00, 1.00}

\newcommand{\head}[1]{{\noindent\textbf{#1.}\xspace}} %
\newcommand{\fig}[1]{{Figure~#1}\xspace} %
\newcommand*\circled[1]{\tikz[baseline=(char.base)]{
           \node[shape=circle,draw,inner sep=0pt,fill=black, text=white] (char) {#1};}}

\definecolor{dpink}{rgb}{0.75, 0.0, 0.75}

\definecolor{amber}{rgb}{1.0, 0.49, 0.0}
\definecolor{awesome}{rgb}{1.0, 0.13, 0.32}
\definecolor{dollarbill}{rgb}{0.52,0.73,0.4}
\definecolor{moegi}{rgb}{0.357, 0.537, 0.188}
\definecolor{burgundy}{rgb}{0.5, 0.0, 0.13}
\definecolor{ballblue}{rgb}{0.13, 0.67, 0.8}
\definecolor{ups-truck}{rgb}{0.53, 0.28, 0.21}
\definecolor{airforceblue}{rgb}{0.36, 0.54, 0.66}
\definecolor{cadmiumgreen}{rgb}{0.0, 0.42, 0.24}
\definecolor{darkcyan}{rgb}{0.0, 0.55, 0.55}
\definecolor{caribbeangreen}{rgb}{0.0, 0.8, 0.6}
\definecolor{flamingopink}{rgb}{0.99, 0.56, 0.67}
\definecolor{jazzberryjam}{rgb}{0.65, 0.04, 0.37}
\definecolor{mediumpersianblue}{rgb}{0.0, 0.4, 0.65}
\definecolor{coolblack}{rgb}{0.0, 0.18, 0.39}
\definecolor{bleudefrance}{rgb}{0.19, 0.55, 0.91}
\definecolor{ao}{rgb}{0.0, 0.0, 1.0}
\definecolor{babyblueeyes}{rgb}{0.63, 0.79, 0.95}
\definecolor{darkwarmgray}{rgb}{0.2,0,0}

\definecolor{dkgreen}{rgb}{0,0.6,0}
\definecolor{gray}{rgb}{0.5,0.5,0.5}
\definecolor{mauve}{rgb}{0.58,0,0.82}

\newcommand{\squishenumstart}[1][1)]{
 \begin{enumerate}[#1]
  { \setlength{\itemsep}{0pt}
     \setlength{\parsep}{0pt}
     \setlength{\topsep}{3pt}
     \setlength{\partopsep}{0pt}
     \setlength{\leftmargin}{1em}
     \setlength{\labelwidth}{1em}
     \setlength{\labelsep}{0.5em} } }

\newcommand{\squishenumend}{
  \end{enumerate}  }

\newcommand{\squishlist}{
 \begin{list}{$\circ$}
  { \setlength{\itemsep}{0pt}
     \setlength{\parsep}{0pt}
     \setlength{\topsep}{3pt}
     \setlength{\partopsep}{0pt}
     \setlength{\leftmargin}{1em}
     \setlength{\labelwidth}{1em}
     \setlength{\labelsep}{0.5em} } }

\newcommand{\squishend}{
  \end{list}  }

\makeatletter
\g@addto@macro{\normalsize}{%
  \setlength{\abovedisplayskip}{3pt plus 0.5pt minus 1pt}
  \setlength{\belowdisplayskip}{3pt plus 0.5pt minus 1pt}
  \setlength{\abovedisplayshortskip}{0pt}
  \setlength{\belowdisplayshortskip}{0pt}
  \setlength{\intextsep}{4pt plus 1pt minus 1pt}
  \setlength{\textfloatsep}{4pt plus 1pt minus 1pt}
  \setlength{\skip\footins}{5pt plus 1pt minus 1pt}}
\makeatother

\titlespacing\section{0pt}{2pt plus 1pt minus 1pt}{3pt plus 1pt minus 2pt}
\titlespacing\subsection{0pt}{2pt plus 1pt minus 1pt}{3pt plus 1pt minus 2pt}
\titlespacing\subsubsection{0pt}{2pt plus 1pt minus 1pt}{3pt plus 1pt minus 2pt}

\newcolumntype{L}[1]{>{\raggedright\let\newline\\\arraybackslash\hspace{0pt}}m{#1}}
\newcolumntype{C}[1]{>{\centering\let\newline\\\arraybackslash\hspace{0pt}}m{#1}}
\newcolumntype{R}[1]{>{\raggedleft\let\newline\\\arraybackslash\hspace{0pt}}m{#1}}

\emergencystretch=\maxdimen
\definecolor{amber}{rgb}{1.0, 0.49, 0.0}
\definecolor{darkamber}{rgb}{0.9, 0.49, 0.0}
\definecolor{darkgreen}{rgb}{0.0, 0.2, 0.13}
\definecolor{darkbyzantium}{rgb}{0.36, 0.22, 0.33}
\definecolor{darkseagreen}{rgb}{0.56, 0.74, 0.56}
\definecolor{darkspringgreen}{rgb}{0.09, 0.45, 0.27}
\definecolor{dollarbill}{rgb}{0.52, 0.73, 0.4}
\definecolor{darkcerulean}{rgb}{0.03, 0.27, 0.49}
\definecolor{poop}{rgb}{0.75, 0.33, 0.0}

\bstctlcite{IEEEexample:BSTcontrol}

\newcommand{\ethz}{{\large$^\dagger$}} 
\newcommand{\cmu}{{\large$^\ddagger$}}

\pdfoutput=1

\title{\myName: \myTitle}

\author[\ethz]{Aditya Manglik}
\author[\ethz]{Minesh Patel}
\author[\ethz]{Haiyu Mao}
\author[]{Behzad Salami}
\author[\ethz]{Jisung Park}
\author[\ethz\cmu]{Lois Orosa}
\author[\ethz]{Onur Mutlu}
\affil[\ethz]{ETH Zürich}
\affil[\cmu]{Galicia Supercomputing Center (CESGA)}

\begin{document}
\maketitle
\thispagestyle{plain}
\pagestyle{plain}

\begin{abstract}

Resistive Random-Access Memory (RRAM) is well-suited to accelerate neural network (NN) workloads as RRAM-based Processing-in-Memory (PIM) architectures natively support highly-parallel multiply-accumulate (MAC) operations that form the backbone of most NN workloads. Unfortunately, NN workloads such as transformers require support for non-MAC operations (e.g., softmax) that RRAM cannot provide natively. Consequently, state-of-the-art works either integrate additional digital logic circuits to support the non-MAC operations or offload the non-MAC operations to CPU/GPU, resulting in significant performance and energy efficiency overheads.

In this work, we propose NEON, a novel compiler optimization to enable the end-to-end execution of the NN workload in RRAM. The key idea of NEON is to transform each non-MAC operation into a lightweight yet highly-accurate neural network. Utilizing neural networks to approximate the non-MAC operations provides two advantages: 1) We can exploit the key strength of RRAM, i.e., highly-parallel MAC operation, to flexibly and efficiently execute non-MAC operations in memory. 2) We can simplify RRAM’s microarchitecture by eliminating the additional digital logic circuits while reducing the data movement overheads. Acceleration of the non-MAC operations in memory enables NEON to achieve a 2.28x speedup compared to an idealized digital logic-based RRAM. We analyze the trade-offs associated with the transformation and demonstrate feasible use cases for NEON across different substrates.

\end{abstract}
\section{Introduction}

Data movement between memory and computation units inhibits the performance of memory-intensive workloads~\cite{mutlu2019processing,sebastian2020memory}. Processing-in-Memory (PIM) offers a potential solution to improve memory-intensive workloads' performance and energy efficiency by reducing the data movement~\cite{song2018graphr,ahn2015scalable,Boroumand2021MitigatingEM,Boroumand2018GoogleWF,ghose2019processing,kwon2018beyond}.
Among different PIM substrates, Resistive Random-Access Memory (RRAM) is under active investigation for accelerating neural network workloads~\cite{wong2012metal,Wong2010RecentPO,li2017resistive,tsai2018recent,kwon2019disaggregated}. 
RRAM's subarrays are composed of \emph{resistive crossbars} that offer in-memory Multiply-ACcumulate (MAC) computation capability~\cite{hosoi2006high,Prezioso2015TrainingAO,Sheu2011A4E,Govoreanu20111010nm2HC}. Prior works utilize this capability to propose RRAM-based neural network accelerators and demonstrate orders of magnitude higher performance and energy efficiency compared to CPU, GPU, and ASICs~\cite{chi2016prime,song2017pipelayer,ankit2019puma}.

\begin{figure}[t]
    \centering
    \includegraphics[width=0.5\linewidth]{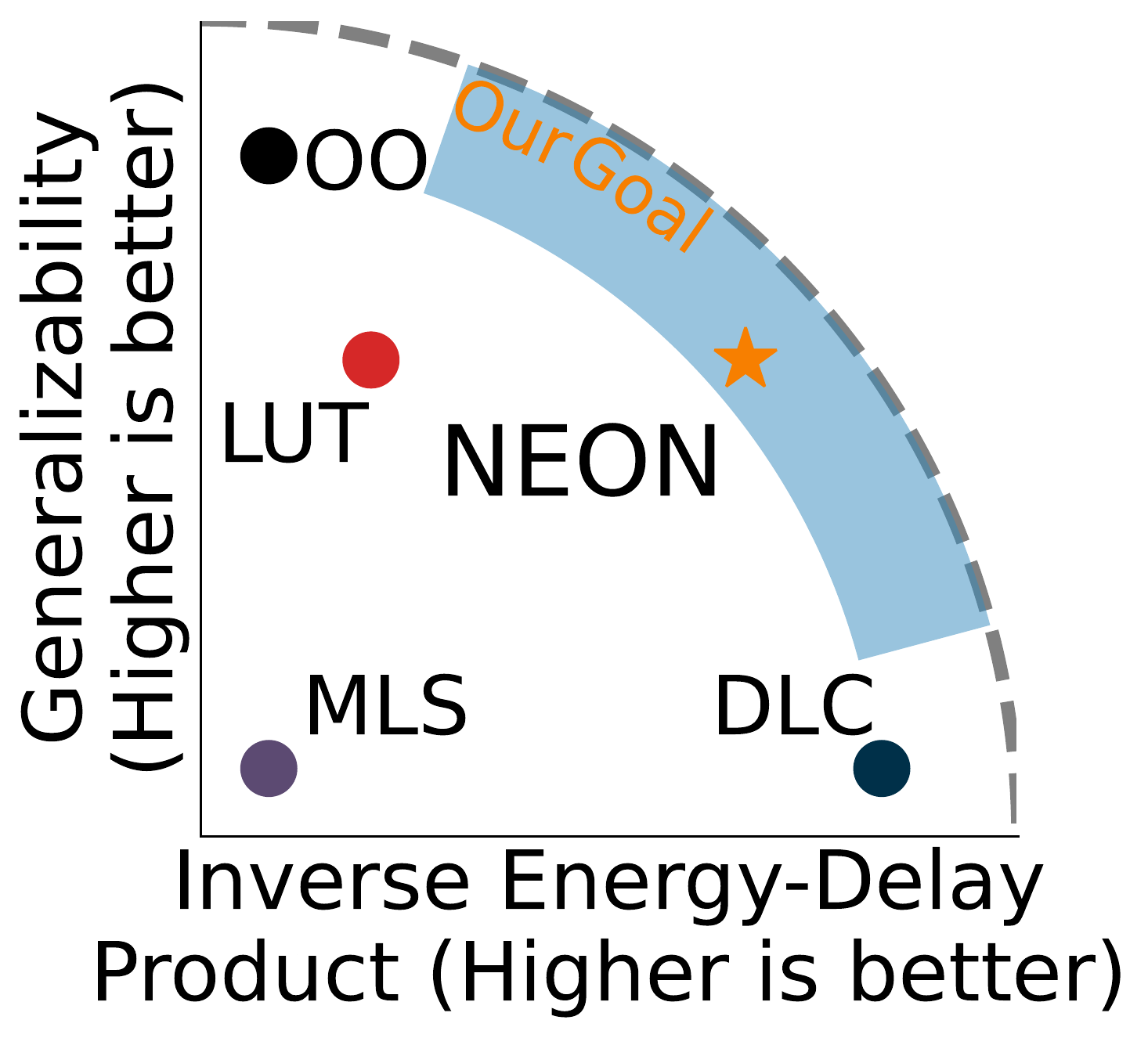}
    \caption{Generalizability vs. $EDP^{-1}$ trade-off for different methodologies to support  non-MAC operations in RRAM\\(OO: Operation Offloading, LUT: Lookup Tables, DLC: Digital Logic Circuits, and MLS: Memristor-based Logic Synthesis).}
    \label{fig:pendulum}
\end{figure}

We survey prior proposals for RRAM-based neural network inference accelerators~\cite{chi2016prime,song2017pipelayer,shafiee2016isaac, Li2015RRAMBasedAA,ji2019fpsa, cheng2017time, Ankit2020PANTHERAP,li2013memristor,mao2018lergan, ji2018recom, wang2020reboc, wang2018snrram, Gupta2019NNPIMAP, lin2019learning, zheng2020lattice,Zhang2021AUR,long2018reram,Liu2015RENOAH, chen2019zara, Chen2018ReGANAP, chen2019efficient, imani2020deep, Bojnordi2016MemristiveBM,ankit2019puma,kim2015reconfigurable,Gupta2018FELIXFA,yang2020retransformer,han2019era, Kenzhina2018AnalysisOH,3dmemristor, Chowdhury2018MBCNNMB,fan2019red, li2020hitm, yan2018celia, yan2019icelia,Marinella2018MultiscaleCA, qiao2018atomlayer, Park2013NeuromorphicSS, Lin2018DemonstrationOG} 
and observe a common design pattern: resistive crossbars execute the MAC operations, and additional computation structures execute the non-MAC operations in the workload. Resistive crossbars cannot execute the  non-MAC operations in neural networks, for instance, softmax, sigmoid, and ReLu~\cite{hayou2019impact}.  Consequently, additional computation structures are integrated into the microarchitecture to support the non-MAC operations. The computation structures are implemented via different methodologies, including digital logic circuits (DLC), lookup tables (LUT), memristor-based logic synthesis (MLS), and operation offloading (OO). Each methodology offers different trade-offs with respect to the ability to support different operations (generalizability) and performance (see Figure~\ref{fig:pendulum}). 

We observe a fundamental restriction of the common design pattern followed by prior proposals: executing different neural network workloads in a single RRAM microarchitecture is difficult and inefficient as the system designer must integrate multiple computation structures in the microarchitecture to support different non-MAC operations in different workloads. 
Further, integrating memory and high-performance logic in a single chip presents manufacturing difficulties and might result in lower yields~\cite{Li2015VariationawareRD,Wong2010RecentPO,Wu2017DeviceAC,Zahoor2020ResistiveRA, Levisse2018RRAMCA}.

\textbf{Our goal} in this work is to enable efficient and generalizable support for different  non-MAC (nonlinear) operations in RRAM-based neural network accelerators. To achieve this goal, we propose \emph{\myName}, (\underline{N}onlin\underline{E}ar \underline{O}peration emulatio\underline{N}), a novel hardware/software co-design methodology that leverages the  resistive crossbars to enable in-memory support for nonlinear operations. \myName is based on three key insights: (1) resistive crossbars offer in-memory MAC execution capability, (2) we can leverage the subarray-level parallelism in RRAM to execute multiple neural networks in parallel, and (3) neural networks can accurately emulate  different operations via the \emph{universal function approximation} theorem~\cite{hornik1993some,Ferrari2005SmoothFA, Petersen2018OptimalAO}. \myName leverages these insights to transform the unsupported nonlinear operations into lightweight neural networks and execute them in RRAM.

Our methodology comprises three components: (1) an automated transformation process for replacing the unsupported operations  in the execution graph with neural networks referred to as \emph{\myName-Nets} (Section~\ref{sec:transformationmechanism}), (2)a simplified RRAM microarchitecture supporting MAC operations (via resistive crossbars) and a single nonlinear operation implemented via DLC. (Section~\ref{sec:microarchitecturemodifications}), and (3) compiler and run-time modifications to effectively integrate the \myName-Net and the \guestName neural network, along with system optimizations. These three components enable a single low-cost RRAM microarchitecture to execute a wide range of  neural networks \emph{flexibly}.

We demonstrate the generalizability of \myName by generating and training \myName-Nets for multiple operations (Table~\ref{tab:neonnetalgorithmperformance})  collected from our evaluation benchmark composed of diverse neural networks  (Section~\ref{sec:systemperfevaluation}). Further, we explore the trade-off space between the performance and accuracy of the transformation process (Section~\ref{sec:epsilonvariation}). We develop an automated tool to implement the transformation process and enable exploration of the \myName-Net design space. We will open source the source code of the transformation tool and the trained \myName-Nets in the final version of the manuscript.

This work makes the following \textbf{contributions}:
\begin{enumerate}
    \item We propose a novel hardware/software co-design methodology for supporting different nonlinear operations in RRAM, \textbf{\myName}. NEON enables a single RRAM microarchitecture to efficiently support in-memory execution for different neural networks. 
    \item We explore the trade-off space between high-performance and high-accuracy \myName-Nets. We will open-source the tool's code to explore the trade-off space and the trained \myName-Nets in the final version of the manuscript.
    \item \myName improves the end-to-end system performance by $2.28\times$ and $1.4\times$ over the DLC and LUT methodologies, respectively. \myName incurs $1.42\times$ higher and $1.17\times$ lower area utilization, $2.02\times$ higher and $1.16\times$ lower power dissipation overheads compared to the DLC and LUT methodologies, respectively.
\end{enumerate}

\section{Background}
We briefly introduce neural networks, followed by RRAM's organization and operation mechanism for accelerating neural network inference. Next, we describe the universal function approximation theorem and prove the generalizability of the \myName methodology based on the theorem.

\subsection{Neural Networks}
\label{sec:bg_neuralnetworks}
Neural networks have emerged as an important class of workloads for applications such as self-driving cars~\cite{bojarski2016end} (using CNNs) and machine translation~\cite{bahdanau2014neural, sennrich2015neural} (using transformers). The programmer defines the neural network's structure before compilation, represented as a Directed Acyclic Graph (DAG). In the DAG, the vertices represent the operations, and the edges represent the data flow between the operations. The compiler can optimize the execution graph before deployment on the target hardware~\cite{ovtcharov2015accelerating, Chetlur2014cuDNNEP}. 

\begin{figure}[h]
        \centering
        \includegraphics[width=1\linewidth]{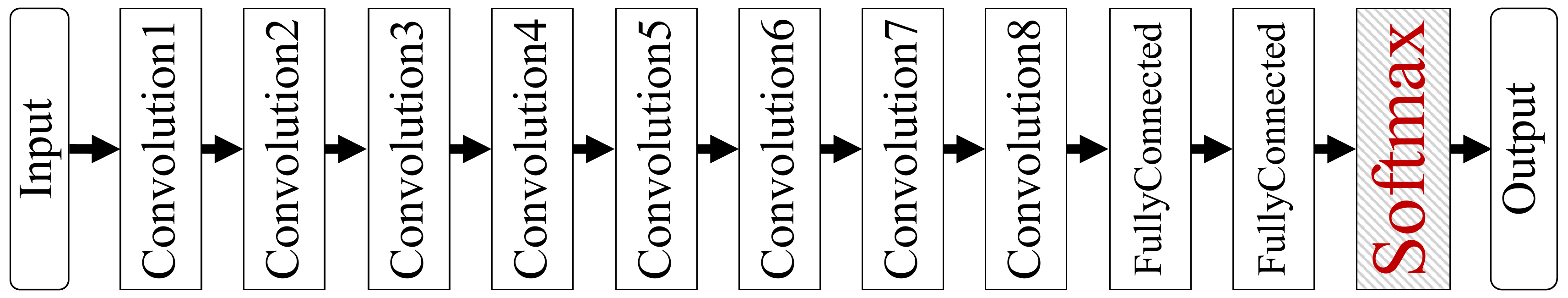}
        \caption{Execution graph for VGG-11~\cite{simonyan2014very} (2014).}
        \label{Fig:neural_network}
\end{figure}

Convolutional Neural Networks (CNNs) are composed of convolutional and fully connected layers (composed of MAC operations).
A convolutional layer comprises several kernels, and the size of each layer is a function of the kernel size and the number of input and output channels. The number of input channels is equal to the depth of the input feature maps (e.g., three for an input layer operating on RGB images). The number of output channels equals the depth of the output feature maps. Convolutional layers are followed by fully-connected layers with a softmax (nonlinear operation) at the output layer (Fig.~\ref{Fig:neural_network}). Prior works exploit this observation to reduce the number of operations supported in the hardware\footnote{Softmax is a common function used by almost all neural networks. Interestingly, only one proposal~\cite{yang2020retransformer} out of the 39 considered in our survey reported in-memory execution capability for softmax.} by offloading softmax operations from the accelerator to the CPU~\cite{Chowdhury2018MBCNNMB, fan2019red}.

\begin{figure}[h]
\centering
    \includegraphics[width = \linewidth]{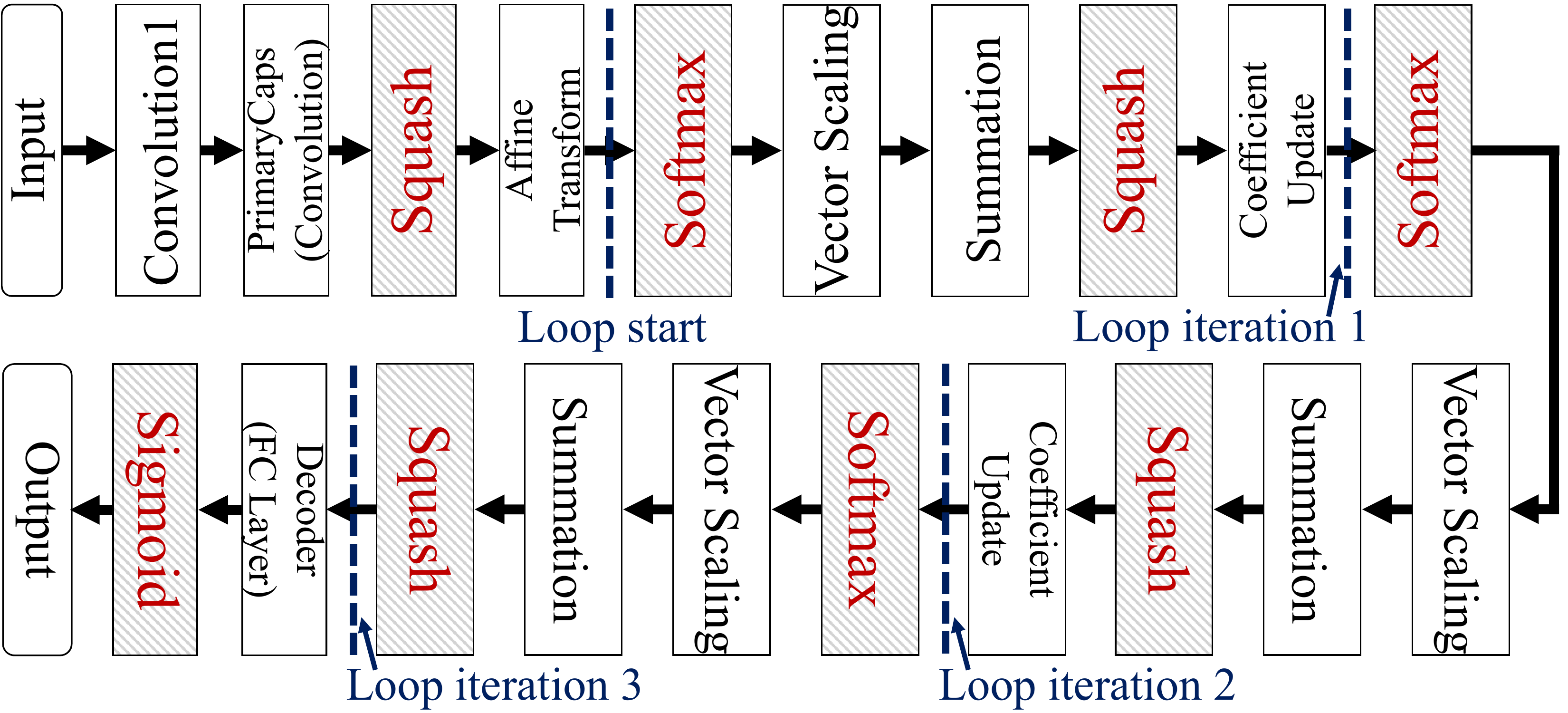}
    \caption{Execution graph for capsule network (2018). Nonlinear operations not supported in RRAM are highlighted in red.}
    \label{fig:capsulediagram}
\end{figure}

\head{Modern neural networks}
\label{sec:bg_modernneuralnetworks}
\fig{\ref{fig:capsulediagram}} shows the execution graph (DAG) for a modern neural network, the capsule network \emph{CapsNet}~\cite{sabour2017dynamic}. We observe squash, softmax, and sigmoid operations in the middle of the critical path. It is difficult to offload these operations to the CPU without significant performance and energy consumption overheads stemming from off-chip data movement and frequent synchronization requirements~\cite{boroumand2019conda, Boroumand2018GoogleWF}.

\subsection{RRAM}
\label{bg:rram_array}
\begin{figure}
    \centering
    \includegraphics[width=\linewidth]{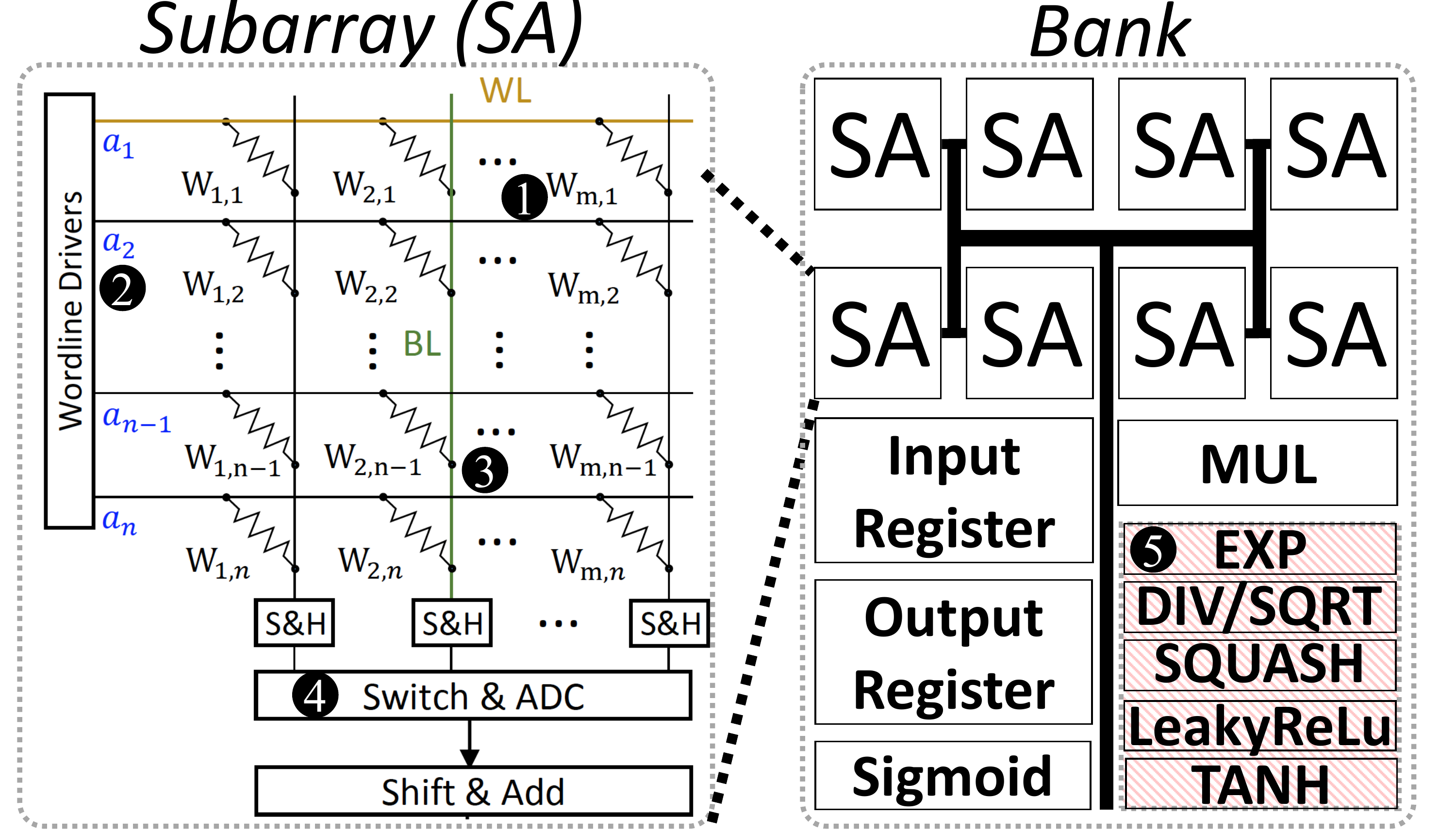}
    \caption{A representative RRAM microarchitecture based on the common design pattern: resistive crossbars execute the MAC operations, and DLCs execute the non-MAC operations.}
    \label{fig:crossbar_diagram}
\end{figure}

\head{Organization}
\fig{\ref{fig:crossbar_diagram}} shows an organizational overview of a reference RRAM microarchitecture designed for accelerating neural network inference. The high-level chip organization comprises multiple memory banks~\cite{chi2016prime}. Each memory bank comprises multiple subarrays and sensing circuitry for performing the memory read and write operations. The one-transistor-one-resistor (1T1R) crossbar structure is used for the subarray microarchitecture due to higher cell selectivity and low leakage current~\cite{Walczyk2011ImpactOT, xu2015overcoming, xia2016technological}. The subarrays are connected via H-tree interconnects~\cite{shafiee2016isaac}. Each subarray can independently execute different operations, resulting in a PIM substrate with massively parallel computation capabilities~\cite{hajinazar2020simdram}.

\head{Operation Mechanism}
We use Figure~\ref{fig:crossbar_diagram} to illustrate the Processing-in-Memory (PIM) operation mechanism of the resistive crossbars. The objective is to execute a vector-matrix multiplication (VMM) between the vector $A$ and the matrix $W$. The matrix $W$ is stored in memory (resistive crossbars), and the vector $A$ is input via the wordlines (WL). First, each value $W_{i,j}$ of the matrix $W$ is encoded as the conductance $G$ (inverse of the cell's resistance $R$, $G = \frac{1}{R}$)~\cite{hu2016dot}. Next, the conductance values $G_{i,j}$ are written to the resistive crossbars based on device characteristics~\cite{zhang2016mellow,zhang2019design,liu2015vortex}. This completes the initialization of the crossbars before the execution (step \circled{1} in Figure~\ref{fig:crossbar_diagram}). 

During execution, each input value in $A$, is converted to an analog voltage value $a_{i}$ through the digital-to-analog converters (DACs). The DACs drive the voltage on the respective wordline (step \circled{2} in Figure~\ref{fig:crossbar_diagram}). Third, the voltage difference drives a current $I_{i} = a_{i} \cdot G_{i,j}$ (inverse Ohm's law; $I=V/R$). The current across the bitline is accumulated to yield the result of the MAC operation between the values stored in the cells on the bitline and the input voltage values (step \circled{3} in Figure~\ref{fig:crossbar_diagram}). Fourth, the accumulated current value is stored in the sample-and-hold circuits (S\&H) and digitized via analog-to-digital converters (ADCs) (step \circled{4} in Figure~\ref{fig:crossbar_diagram}). Fifth, the digitized values are forwarded to the additional computation structures for executing the nonlinear operations (step \circled{5} in Figure~\ref{fig:crossbar_diagram}).

\begin{figure}[h]
\centering
    \includegraphics[width = \linewidth]{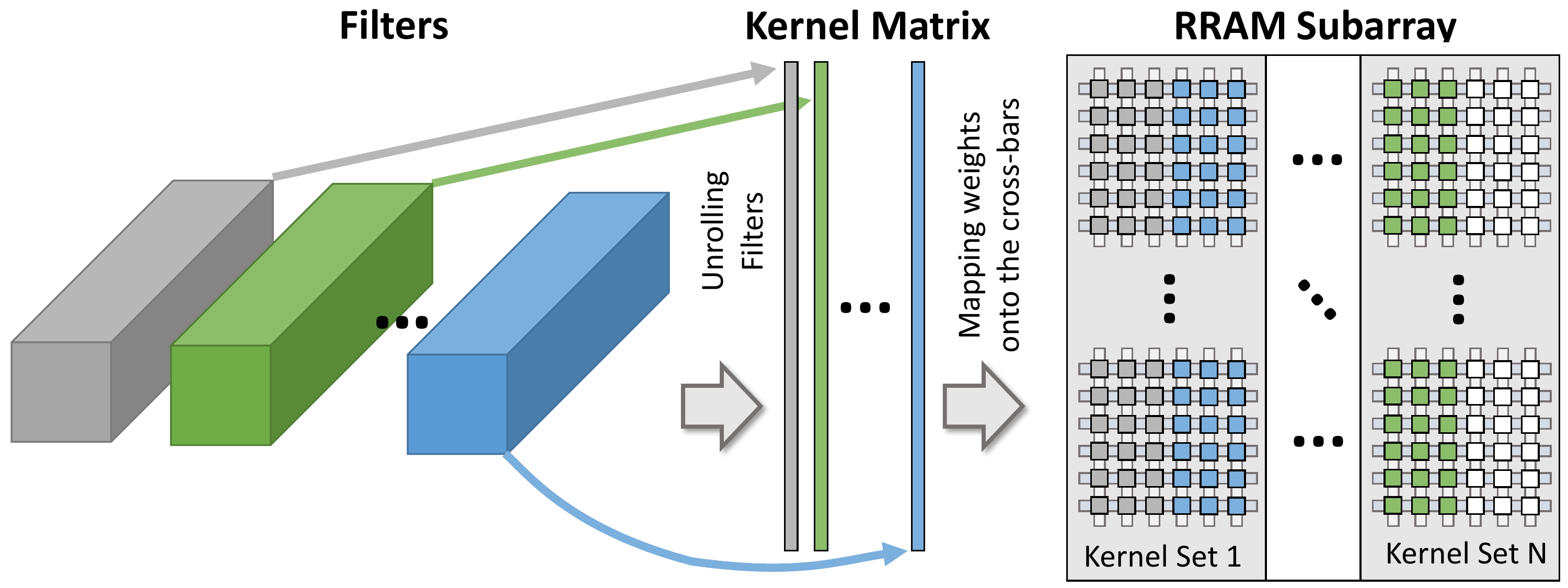}
    \caption{Mapping a CNN to the RRAM Subarray comprising resistive crossbars and additional computation structures.}
    \label{fig:rrammapping}
\end{figure}

\head{Mapping Neural Network Workloads on RRAM} 
\label{bg:mappinginrram}
To map the workload's execution graph (DAG) on the resistive crossbars, each layer's weight matrices are unrolled depth-wise and represented as a vertical column, referred to as a kernel. Every output channel in the layer is considered a single kernel, as shown in Fig.~\ref{fig:rrammapping}. Kernels may be split or duplicated across different RRAM subarrays~\cite{wang2019deep, zhang2020efficient} based on the size and the number of input and output channels. Kernel sizes can significantly influence the \emph{utilization ratio} for resistive crossbars. As an illustrative example, kernel set N in Figure~\ref{fig:rrammapping} utilizes only half of the available subarray capacity, resulting in a 0.5 utilization ratio. High utilization ratios (maximum 1.0) are desirable for higher energy efficiency~\cite{hanif2020resistive}.

\head{Precision}\label{sec:precision}
To execute an n-bit fixed-point MAC operation between the mapped weights and the input value $A_i$, 1-bit DACs inject n bits successively over n input cycles~\cite{shafiee2016isaac}. We use separate subarrays for mapping the positive and negative weight values. Multi-bit weight values are distributed across different columns due to the limited precision of a single memristor cell~\cite{lin2018rescuing, Zangeneh2014DesignAO} (observe multiple columns for each kernel in Fig.~\ref{fig:rrammapping}). The final results are summed across different columns via Shift-and-Add (S\&A) units.

\subsection{Universal Function Approximation (UFA) theorem}
\label{sec:universalapproximationtheorem}
The universal function approximation theorem (UFA) states that a feed-forward neural network with at least one hidden layer and a continuous\footnote{ReLu is a notable exception. The function is not continuously differentiable, but the gradient is defined for the discontinuity as a special case~\cite{daubechies2022nonlinear}.}, bounded and non-constant activation function can approximate a function boundary with arbitrary precision~\cite{Cybenko1989ApproximationBS, hornik1991approximation, hornik1993some,chen1995universal, huang2006universal, Ito1994ApproximationCO}. Prior work has used the UFA theorem to approximate different operations in the C mathematical library with neural networks~\cite{Eldridge2014NeuralNA}, and replace code regions in general-purpose workloads with neural networks~\cite{esmaeilzadeh2012neural}.

\section{Trade-offs of Different Methodologies for Supporting Operations in RRAM}
\label{sec:problem}

This section describes the results of our survey of prior proposals for accelerating neural networks in RRAM, followed by the trade-offs associated with each methodology in the survey.%

\head{Survey}
We survey prior RRAM-based neural network inference accelerators and categorize them based on the methodology for supporting nonlinear operations. Table~\ref{tab:function_lit_survey} reports the survey results for each methodology: Digital Logic Circuits (DLC), Lookup Tables (LUT), memristor-based logic synthesis (MLS), offloading operations (OO) to the host, and no discussion.

\begin{table}[h!]
\begin{center}
 \resizebox{0.95\linewidth}{!}
 {
 \begin{tabular}{ccc}
 \toprule
         \textbf{Prior work (Count)} & \textbf{Methodology} & \textbf{Nonlinear Operations} \\ \midrule
         \makecell{\cite{chi2016prime, Li2015RRAMBasedAA, shafiee2016isaac, ji2019fpsa, cheng2017time, Ankit2020PANTHERAP,li2013memristor,mao2018lergan, ji2018recom, wang2020reboc, wang2018snrram, Gupta2019NNPIMAP, lin2019learning, zheng2020lattice,Zhang2021AUR,long2018reram}} (16) & \makecell{Digital Logic\\Circuits (DLC)} & Sigmoid, ReLu \\ \midrule
         \cite{Liu2015RENOAH, chen2019zara, Chen2018ReGANAP, chen2019efficient, imani2020deep, song2017pipelayer, Bojnordi2016MemristiveBM,ankit2019puma,kim2015reconfigurable} 
         (9) & Lookup tables (LUT) & \makecell{Sigmoid, ReLu,\\LeakyReLu} \\ \midrule
         \cite{Gupta2018FELIXFA,yang2020retransformer,han2019era, Kenzhina2018AnalysisOH}
         (4) & \makecell{Memristor-based\\Logic Synthesis (MLS)} & \makecell{Softmax, Sigmoid\\tanh} \\ \midrule
         \cite{3dmemristor, Chowdhury2018MBCNNMB} (2) & \makecell{Operation\\Offloading (OO)} & \makecell{Squash, Softmax,\\ Sigmoid, ReLu} \\ \midrule
         \makecell{\cite{fan2019red, li2020hitm, yan2018celia, yan2019icelia,Marinella2018MultiscaleCA, qiao2018atomlayer, Park2013NeuromorphicSS, Lin2018DemonstrationOG}} (8) & \makecell{Nonlinear Operation\\support not discussed} & - \\
        \bottomrule
 \end{tabular}
 }
\caption{Nonlinear operation survey in RRAM}
\label{tab:function_lit_survey}
\end{center}
\end{table}
\noindent Next, we discuss the trade-offs for each methodology.
\subsection{Digital Logic Circuits (DLC)}
\label{sec:digitallogicapproach}
16 out of 39 works in our survey support nonlinear operations via the integration of DLC in the microarchitecture. DLCs offer low latency and low area requirements but suffer from two drawbacks:
1) Fixed-function circuits restrict the microarchitecture to a limited set of nonlinear operations. The operations must be known at design time and cannot be changed after the chip is manufactured. %
Flexible logic units (e.g., Chebyshev~\cite{oliveira1973generalised} and Taylor series-based function approximation~\cite{pang2010optimization,parhi2016computing}) offer marginally higher generalizability but result in significantly worse performance (due to the use of multiplications to calculate the output value) and a larger area requirement compared to fixed-function circuits~\cite{long2018reram}. Notably, only one prior work~\cite{long2018reram} in our survey uses flexible logic units.
2) Power dissipation is a key constraint for PIM substrates~\cite{kim2018practical,Jagasivamani2019DesignFR,Xue2019241A1}. Static power dissipation restricts the number of DLCs that may be integrated in the microarchitecture. As an example, supporting 1152-dimensional softmax in CapsNet requires $576\times$ \texttt{EXP} and \texttt{DIV} units that incur enormous static power overhead (19.76 W).

\subsection{Lookup Tables (LUTs)}
\label{sec:lookuptablelimitation}
9 out of 39 works in our survey utilize LUTs to support nonlinear operations. LUTs offer flexibility but suffer from two drawbacks:
1) Large memory requirement restricts the scalability of LUTs. Scalability is required to support parallel nonlinear operations on the critical path. For instance, nonlinear operations in CapsNet require 23040 LUTs that consume 2880 MB, 424$\times$ larger than the memory requirement of CapsNet's weights~\cite{marchisio2019capsacc}.
2) The size of the RRAM subarray is generally restricted (e.g., 128 or 256-sized crossbar) to improve switching capability and minimize noise effects~\cite{xia2016technological, Liu2014ReductionAI,bhattacharjee2022examining}. Consequently, large LUTs must be divided across several subarrays, leading to hierarchical memory accesses that incur significant latency penalties~\cite{zhang2003incrementally}.

\subsection{Memristor-Based Logic Synthesis (MLS)}
\label{sec:MAGIClimitation}
MAGIC~\cite{kvatinsky2014magic, haj2018not, talati2016logic, hur2017simple} proposes synthesizing primitive logic operations such as \texttt{XOR} and \texttt{NOR} using memristor cells. 4 out of 39 works leverage MLS to support nonlinear operations in RRAM. MLS is unable to offer either performance or generalizability due to the following three drawbacks:
1) Memristors exhibit asymmetric read/write performance, and write operations consume three orders of magnitude higher energy~\cite{niu2013design,zhang2016mellow} than read operations. Each MAGIC-based logic gate requires 2-3 memory write operations per input that incurs significant energy consumption overheads.
2) The system designer must combine primitive operations such as \texttt{NOR} into complex digital operations. For instance, more than 20,000 memristor cells are needed to implement a simple 16-bit multiplier unit that must be further combined into complex operations such as \texttt{EXP} in softmax. 
3) The synthesized logic is not reconfigurable and must be fixed at design time, resulting in a fixed-function microarchitecture.

 \subsection{Operation Offloading (OO)} \label{sec:operationoffload}
Two prior works in our survey rely on the host for supporting nonlinear operations via offloading. Theoretically, OO offers generalizability as the host is assumed as a general-purpose CPU. However, OO suffers from two drawbacks: 
1) Frequent communication and synchronization due to multiple calls for nonlinear operations in the middle of the critical path (e.g., \fig{\ref{fig:capsulediagram}}) restrict the accelerator's performance benefits. 
2) Excessive off-chip data movement restricts the accelerator's energy efficiency benefits. For instance, we evaluate the data movement costs between RRAM and CPU using the HyperTransport link used in a prior RRAM accelerator~\cite{shafiee2016isaac}. Data movement stemming from operation offloading requires 64.49$\times$ higher latency and 2.38$\times$ more energy than the latency and energy required for executing the MAC component of the target workload in RRAM.

\section{\myName Methodology}
\label{sec:neonnetdesign}
To enable efficient and generalizable support for nonlinear operations in RRAM-based neural network accelerators, we introduce \myName (\underline{N}onlin\underline{E}ar \underline{O}peration emulatio\underline{N}). \myName is a novel hardware/software co-design methodology to efficiently support different nonlinear operations in RRAM. Next, we describe the key insights, followed by an overview of \myName. Further, we detail each component and its implementation: the transformation process, RRAM microarchitecture, and the compiler support. Finally, we discuss a key feature of \myName-Nets, Operator Scalability.

\head{Key Insights}
\label{sec:keyinsight}
We base our idea on three key insights: 

(1) Resistive crossbars offer in-memory MAC execution capabilities.

(2) We can leverage the subarray-level parallelism in RRAM to execute multiple neural networks in parallel.

(3) Neural networks can accurately emulate nonlinear operations via the \emph{universal function approximation} theorem~\cite{Ferrari2005SmoothFA, Petersen2018OptimalAO, hornik1993some}.

Based on these insights, \myName generates and trains neural networks to replace the unsupported nonlinear operations in the target neural network's execution graph. \myName leverages the inherent strengths of RRAM subarrays (i.e., parallel and energy-efficient MAC execution) to execute the nonlinear operations within the subarray itself. For the rest of the paper, we refer to the target neural network as the \emph{\guestName} and the generated neural networks as \emph{\myName-Nets}. The workload is assumed to be pre-trained and the training dataset is available during the transformation.

\begin{figure}[h]
\centering
  \begin{subfigure}[b]{0.333\textwidth}
    \includegraphics[width=\linewidth]{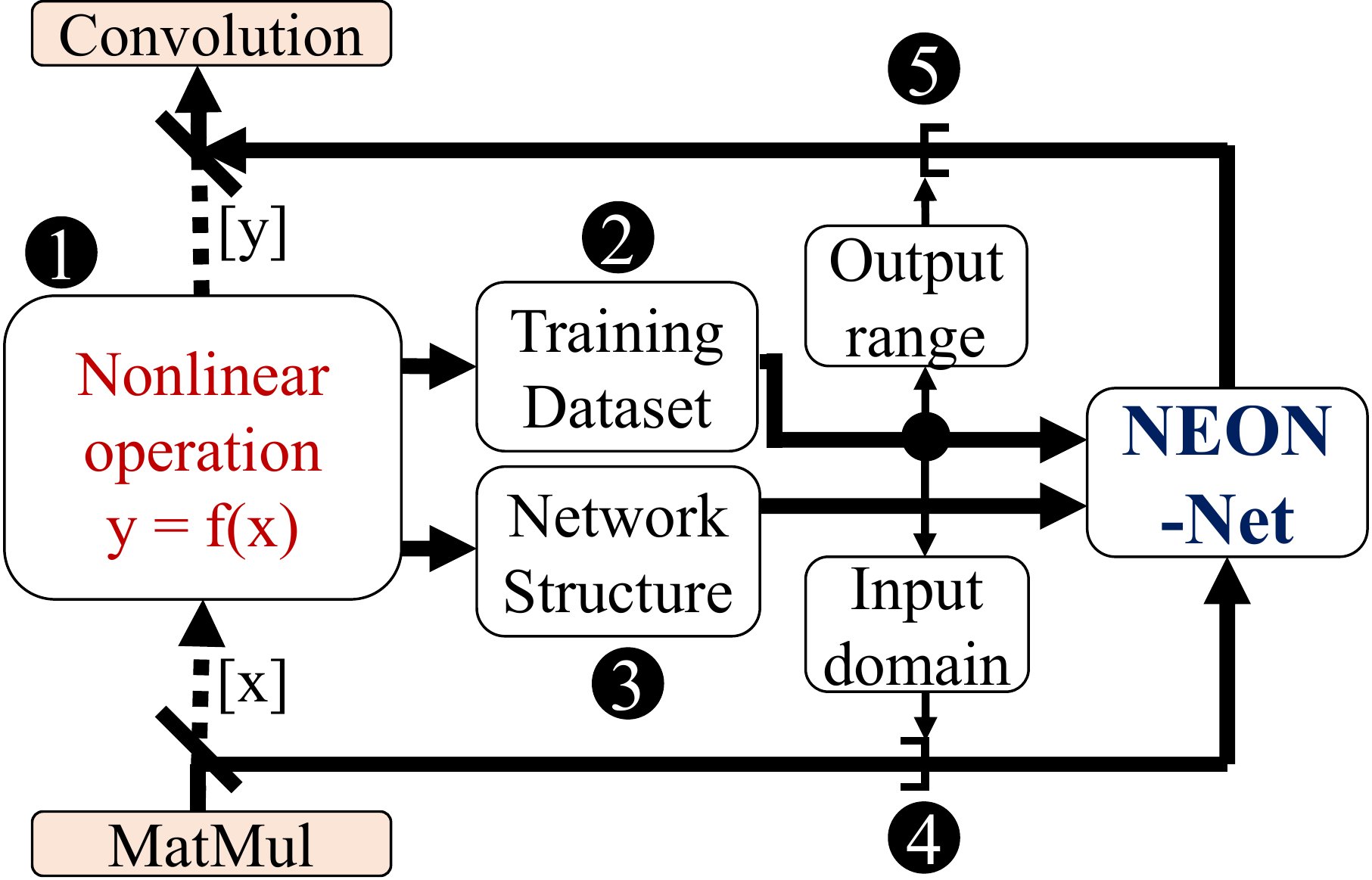}
    \caption{}
    \label{fig:transformationprocess}
  \end{subfigure}%
  \begin{subfigure}[b]{0.166\textwidth}
    \includegraphics[width=\linewidth]{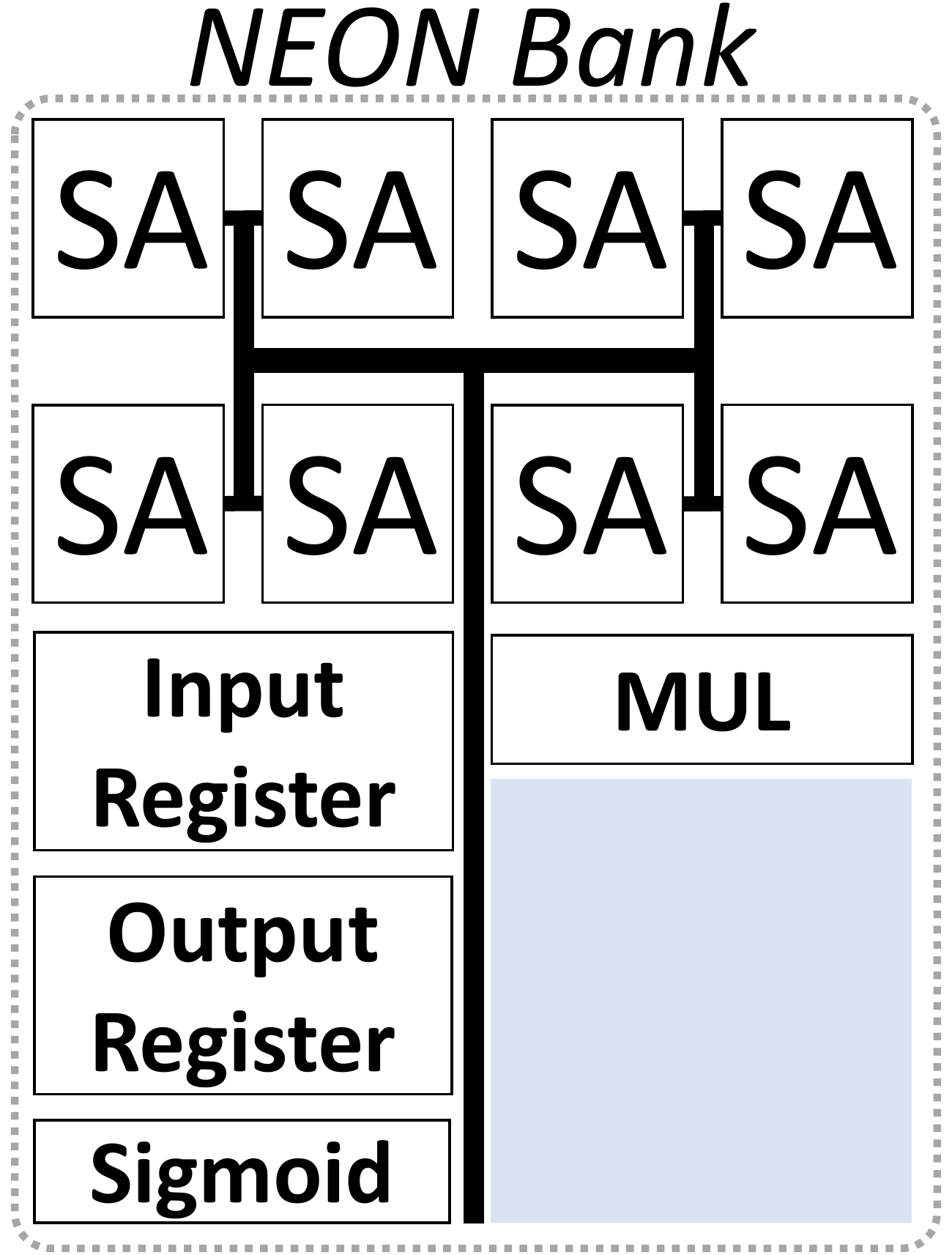}
    \caption{}%
    \label{fig:near_bank}
  \end{subfigure}
  \caption{(a) displays the transformation process, and (b) displays the \myName microarchitecture.}
  \label{fig:NEARFramework}
\end{figure}

\head{Overview} The methodology has three components: (1) an automated \emph{transformation process} for generating and training \myName-Nets (Section~\ref{sec:transformationmechanism}), (2) a simplified \emph{RRAM microarchitecture} capable of executing the transformed workload (Section~\ref{sec:microarchitecturemodifications}), and (3) \emph{compiler} and run-time modifications to effectively integrate the \myName-Net and the \guestName neural network. Figure~\ref{fig:NEARFramework} shows an overview of the methodology. 
Next, we describe each component in detail. 
\subsection{Transformation Process}
\label{sec:transformationmechanism}
\head{Motivation} The UFA theorem guarantees that a neural network can approximate a function boundary with arbitrary precision~\cite{Cybenko1989ApproximationBS,hornik1991approximation, hornik1993some}. However, the theorem does not provide any information about the structure of the neural network or \emph{how to design it}\footnote{Similar limitations apply to other universality theorems, for instance, the universality of Boolean logic~\cite{haastad1987computational}. Also referred to as an existence proof, the theorem guarantees the existence of the solution but does not inform us about how to find the solution.}. The programmer is responsible for determining the \myName-Net's structure. 
This is a non-trivial problem and requires significant expertise in designing and training neural networks~\cite{glorot2010understanding}. To overcome this limitation and make \myName generalizable across different nonlinear operations, we develop an automated transformation process to generate \myName-Nets using information from the workload's execution graph.
Figure~\ref{fig:transformationprocess} illustrates the transformation process comprising three steps:
\begin{enumerate}
    \item Code segment delineation (\circled{1} in Fig.~\ref{fig:NEARFramework})
    \item \myName-Net training dataset generation (\circled{2} in Fig.~\ref{fig:NEARFramework})
    \item \myName-Net structure definition (\circled{3} in Fig.~\ref{fig:NEARFramework})
\end{enumerate}

Next, we detail each step in the transformation process.

\subsubsection{Code Segment Delineation}
\label{sec:identifyingapproximationcandidate}
The first step is identifying the nonlinear operation's code segment for replacement with the \myName-Net. 
Neural network frameworks such as PyTorch and TensorFlow~\cite{abadi2016tensorflow, paszke2017pytorch} are widely used in industry and academia for programming and compiling neural networks. These frameworks expose a stable application programming interface (API) for operations used as common blocks across different neural networks. The operations are implemented via \emph{subroutines}, e.g., softmax subroutine in PyTorch~\cite{PyTorch}. We use the framework's API to identify the subroutine as the code segment for transformation in the workload's execution graph. We fix the granularity of replacement to subroutines.%

\subsubsection{\myName-Net Training Dataset Generation}
The compiler generates the \myName-Net's training dataset using the \guestName and the \guestName's training dataset. To generate the dataset, the compiler executes inference over the \guestName using its training dataset and collects the input and output parameters for the target subroutine ($x$ and $y$ in Listing 1). The input parameter values ($x$) serve as the input feature values, and the output parameter values ($y$) serve as the ground truth in the \myName-Net training dataset.

\subsubsection{\myName-Net Structure Definition}
\label{sec:structuredefinitionNEARNet}

We choose fully-connected (FC) neural networks~\cite{zhang2017learnability} as the base structure for generating \myName-Nets as FC layers are composed of MAC operations that can be directly executed in resistive crossbars. It is possible to consider alternate neural network classes such as CNNs, RNNs, and autoencoders~\cite{burda2015importance} as base structures. However, each class requires co-designing the microarchitecture to ensure end-to-end execution capability and further exploration is left for future work. 

A \myName-Net is a regression neural network composed of at least three FC layers: input, hidden, and output. The compiler uses information from the \guestName's execution graph to determine the input and output layer sizes. Using the softmax subroutine in Listing 1 as an example, the function's input (x) and output (y) parameter's dimensions (d) are used to define the \myName-Net's input and output layer sizes (d nodes in each layer).

\vspace{0.15ex}
\begin{lstlisting}[caption={Softmax subroutine in Python programming language (implementation based on the NumPy library~\cite{harris2020array})},label={algo:softmaxcode},language=Python]
def softmax(x): # x is a d-dimensional vector
    e_x = np.exp(x) # pointwise exponentiation
    y = e_x / e_x.sum() # pointwise division
    return y # y is a d-dimensional vector
\end{lstlisting}

As the input and output layer sizes are fixed, the number and size of hidden layers primarily influence the \myName-Net's performance and accuracy. We measure the accuracy via Mean-Square-Error (MSE)~\cite{allen1971mean} metric (denoted by $\epsilon$). We picked MSE based on the highest end-to-end workload output accuracy across different metrics, including cosine similarity, mean absolute error, and MSE.

\head{Determining the number and size of hidden layers} We develop an algorithm to determine the number and size of hidden layers based on the threshold accuracy indicated by the system designer. The algorithm takes the initial \myName-Net structure with input and output layers and iteratively adds hidden layers until the threshold accuracy is achieved. We constrain the maximum possible number of hidden layers to 100 %
to ensure that the training process completes in a finite period. The number of nodes in the hidden layer is determined based on the target RRAM microarchitecture's crossbar size (e.g., 128 or 256). This allows \myName to maximize the utilization of the subarrays and encourages over-fitting (desirable as the function boundary is deterministic~\cite{Zhang2017UnderstandingDL}).

\begin{algorithm}[h]
    \footnotesize
    \SetAlgoLined
    Input: \myName-Net a single hidden layer, generated training dataset, $\epsilon=10^{-3}$, $max\_layers$ = 100\, $num\_epochs$=100, $XBar\_size$=128;
    initialization: Split dataset into train-validation, counter = 1\;
    \While{$\|f(x)-\epsilon \| > 0$ and $counter<max\_layers$}{
    Train the \myName-Net for $num\_epochs$ using the training dataset;\\
    Determine \myName-Net's MSE ($f(x)$) on validation dataset;\\
    \eIf{$\|f(x)-\epsilon \| < 0$}{
        return trained \myName-Net structure and weights;
    }
    {
    Add a hidden layer with XBar parameters;\\
    Re-initialize the \myName-Net weights;\\
    counter += 1;\\
    }
    }
    \caption{\myName-Net hidden layer structure generation algorithm}
    \label{algo:hiddenlayerdefinition}
\end{algorithm}

\subsection{Microarchitecture Design}
\label{sec:microarchitecturemodifications}
\myName replaces the nonlinear operations with a \myName-Net. The \myName-Net is composed of MACs and a single nonlinear operation. The resistive crossbars directly support the MAC operations. However, the microarchitecture must support the nonlinear operation in the \myName-Nets. We integrate a single fixed-function unit to support the \myName-Net's nonlinear operation. We pick DLC over LUT-based implementation for the implementation as DLCs offer higher performance compared to LUTs for implementing a single function.

Microarchitectural support for the \myName-Net's nonlinear operation ensures that the \myName-Nets themselves are not recursively transformed. Each nonlinear operation in the execution graph is transformed only once, as successive approximations lead to an infinite recursion problem.

\head{Determining the activation function for \myName-Net} The UFA theorem places restrictions on which functions may be used as activation functions. The function must exhibit the following mathematical properties: nonlinear, continuous, and finite output~\cite{hornik1990universal}. Based on these constraints, we evaluate different functions as potential candidates: sigmoid, tanh, and ReLu. We perform a grid search by training a fixed \myName-Net with different activation functions: ReLu (MSE: 0.0774), tanh (MSE: 0.0399), and sigmoid (MSE: 0.3036). We pick tanh as it offers the lowest MSE compared to other activation functions. We constrain all layers in the \myName-Net to use tanh as the common activation function. It might be possible to improve \myName-Net's accuracy with more experiments. However, further exploration is left for future work.

\subsection{Compiler Support}
\label{sec:compilersupport}
We describe how the compiler identifies subroutines for transformation, fine-tunes the post-transformation workload to recover the accuracy loss, and a run-time optimization to improve system stability.

\head{Transformation Candidates} RRAM-supported subroutines (e.g., convolution) are scheduled directly, and RRAM-unsupported subroutines (e.g., softmax) are marked as transformation candidates. Any kernel that is directly supported on RRAM is not replaced. Each nonlinear operation is replaced with a separate \myName-Net trained on the compilation system.

\begin{table*}
\small
  \begin{center}
  \resizebox{0.95\linewidth}{!}{
  \begin{tabular}{ ccccccccccc } 
 \toprule
 \textbf{Neural Network} & \textbf{Application Class} & \textbf{Dataset} & \textbf{Input size} & \textbf{Output size} & \textbf{\makecell{Number of\\Parameters}} & \textbf{Description} & \textbf{\makecell{Nonlinear\\ Operation (Dim)}} & \textbf{\makecell{Output\\Accuracy loss}} & \textbf{\makecell{Fine-tuning\\Time}} \\ \midrule
 \makecell{CNN: \textbf{VGG-16}} & Image Recognition & ImageNet & 224x224x3 & 1000x1 & 138 million & VGG-16 reference model & softmax (1000) &  -1.36\% & 19 min \\ \midrule
 \makecell{GRU-based \textbf{RNN}} & Speech Recognition & \makecell{LibriSpeech\\ASR corpus} & 128x1 & 128x1 & 594, 432 & \makecell{Input vector size = 128\\Hidden-state size=128} & \makecell{sigmoid (1),\\softmax (2048)} & 0.1\% & 6.5 min \\ \midrule
 \makecell{Capsule Network\\\textbf{CapsNet}} & \makecell{Occluded\\Object Detection} & CIFAR-10 & 32x32x3 & 16x10 & 6.81 million & \makecell{3x dynamic routing\\ iterations} & \makecell{squash (8), sigmoid (1)\\softmax (1152)} & -1.8\% & 11 min \\ \midrule
 \makecell{\textbf{Transformer}} & \makecell{Neural Machine\\Translation} & \makecell{WMT 2016\\Translation Task} & 128x$d_{model}$ & 128x$d_{model}$ & 60 million & \makecell{Self-Attention Heads = 8\\Encoder/Decoder blocks = 6\\$d_{model}$=512, $d_{ff}$=2048} & \makecell{SQRT (1), sigmoid (1),\\softmax (64)}  &  0.87\% & 15 min \\ 
 \bottomrule
\end{tabular}
}
\caption{Benchmark neural networks}
\label{tab:networkbenchmark}
\end{center}
\end{table*}

\head{Fine-tuning the \guestName after transformation}
\label{sec:finetuninghost}
The transformation process leads to a slight accuracy loss in the \guestName. We can recover the accuracy loss by fine-tuning the \guestName as neural networks are inherently tolerant to approximate execution~\cite{Rakin2019ParametricNI, koppula2019eden}. Fine-tuning is performed after replacing all nonlinear operations with \myName-Nets.

\textbf{Mechanism:} We assume the \guestName is pre-trained before the transformation, and the original training dataset is assumed to be available for fine-tuning. (1) We freeze (mark the layer as untrainable) all layers in the \guestName after replacing the unsupported operations with \myName-Nets. (2) We unfreeze (mark as trainable) one layer before and after the \myName-Net's position in the \guestName. The \myName-Net's layers are marked as frozen. (3) We resume training for the \guestName using the original training dataset. (4) In the forward propagation step, the unsupported subroutine's output is derived from the corresponding \myName-Net's output for the input values. (5) In the backward propagation step, the unsupported subroutine's output is derived from the original function implementation (assumed as available in the compilation system, e.g., GPU). Our fine-tuning mechanism is similar to training methods for low-precision neural networks~\cite{Zhou2016DoReFaNetTL, Banner2018ScalableMF}).

\head{Run-time} At run-time, the call to the unsupported subroutine is replaced with an equivalent function call to execute inference on the \myName-Net (co-executed in a dedicated subarray along with the \guestName). The function's arguments are used as the input values for inference, and the \myName-Net's output replaces the function's return parameters on the call stack. 

Next, we describe a run-time optimization to improve the system's stability.

\head{Input domain and output range constraints}
\label{sec:inputdomainconstraints}
Mathematically, a function may have an infinite input domain $(-\infty, \infty)$ that cannot be realized in practice. To overcome this problem, DLCs and LUTs exploit mathematical properties of functions such as the periodicity of trigonometric functions~\cite{amin1997piecewise,stineman1980consistently}. To overcome this problem for \myName, we constrain the function's input domain and output range by measuring the expected distribution of values from the \guestName\ at run-time.%

\begin{figure}[h]
\centering
  \begin{subfigure}[b]{0.5\textwidth}
    \includegraphics[width=\linewidth]{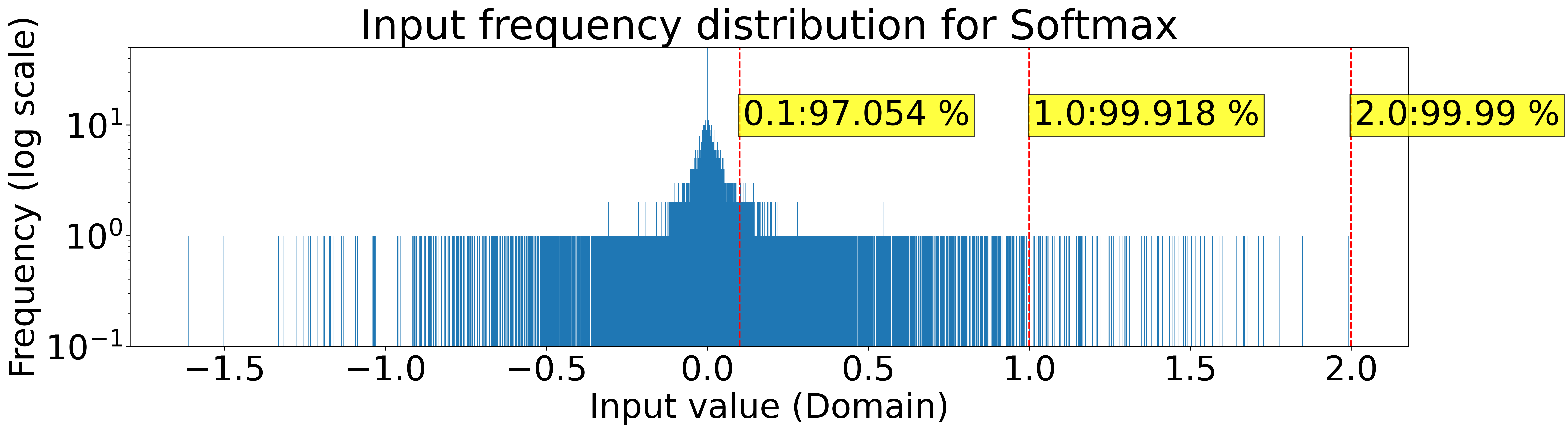}
    \caption{}
    \label{fig:softmaxdistribution}
  \end{subfigure}
  
  \begin{subfigure}[b]{0.5\textwidth}
    \includegraphics[width=\linewidth]{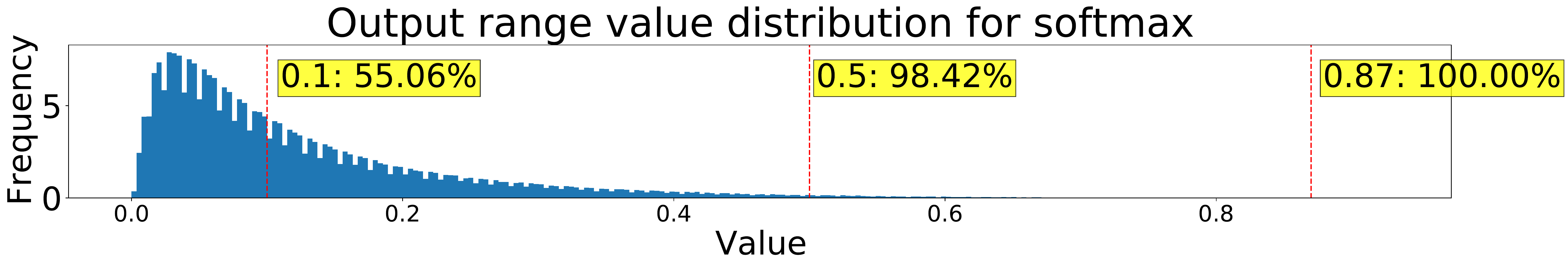}
    \caption{}%
    \label{fig:softmaxoutputrangedistribution}
  \end{subfigure}
  \caption{Softmax input and output value distributions were obtained from CapsNet trained on the CIFAR10 dataset. Extreme values are pruned for representation purposes.}
\end{figure}

Neural networks often use batch normalization~\cite{ioffe2015batch, ba2016layer} after each layer to constrain the values to a normal distribution. For example, Fig.~\ref{fig:softmaxdistribution} illustrates the input value distribution ($x$) for softmax in CapsNet (trained over the CIFAR-10 dataset). We observe a normal distribution with 99.918\% values less than 1.0 and 100\% values less than 9.05. We leverage this observation to constrain the \myName-Net's input domain to $[-1.8, 9.05]$. Similarly, Fig.~\ref{fig:softmaxoutputrangedistribution} illustrates softmax's output value distribution ($y$). We observe that 55\% of the values are less than 0.1, and 100\% are less than 0.87. Output values are constrained to $(0.0, 0.87]$. %

The compiler extracts the expected distribution of input and output values from the \myName-Net training dataset. It parses the input feature and ground truth values to determine the minimum and maximum bounds for the input domain and output range, respectively. 
At run-time, if any value outside these constraints is encountered, it is rounded to the closest value within the bounds (\circled{4}, and \circled{5} in Fig.~\ref{fig:NEARFramework}). This is achieved by adding a simple rounding circuit (consisting of a comparator and a multiplexer).

\subsection{Operator Scalability}\label{sec:inputoperatorscaling}
Digital logic typically accepts unary/binary input values, for example, \texttt{EXP} (unary) and \texttt{DIV} (binary) operations. Increasing the number of input values (operators) incurs a linear increase in the area and power requirements for the digital logic circuits. Similarly, scaling an N-input LUT incurs an exponential increase in the memory requirements ($2^N$). %
In contrast to DLCs and LUTs, \myName-Nets offer a sublinear increase in the Energy-Delay Product (EDP) when scaling the number of operators. The number of input operators for a \myName-Net equals the number of nodes in the input layer. The number of input nodes is directly proportional to the number of wordline activations in the crossbar. 
Consequently, increasing the number of input operators increases the number of wordline activations, resulting in sublinear scaling (until we exceed the capacity of the subarray).

\begin{table*}[h]
\small
 \begin{center}
 \resizebox{0.95\linewidth}{!}
 {
 \begin{tabular}{cccccccc}
 \toprule
 \textbf{Nonlinear Operation} & \textbf{Input Domain} & \textbf{Output Range} & \textbf{Accuracy (MSE)} & \textbf{Hidden layers} & \textbf{Training Time} & \textbf{Area ($mm^2$)} & \textbf{Power (mW)} \\ \midrule
 \textbf{Softmax (64-dim)} & [-3.78, 7.5] & [0.001, 0.87] & $1.5\times10^{-8}$ & 1 & 12.5 min & 0.16 & 288.96 \\
 \textbf{Softmax (1000-dim)} & [-4.08, 5.45] & [0.01, 0.89] & \textbf{$2.62\times10^{-9}$} & 1 & 11.99 min & 1.66 & 3058.16 \\
 \textbf{Softmax (1152-dim)} & [-1.08, 9.05] & [0.01, 0.87] & $2.5\times10^{-5}$ & 2 & 15.57 min & 1.99 & 3660.16 \\
 \textbf{Softmax (2048-dim)} & [-0.08, 6.78] & [0.01, 0.79] & $3.4\times10^{-5}$ & 2 & 18.7 min & 3.45 & 6357.12 \\
 \textbf{Square-root (SQRT)} & [-3.2, 4.5] & [0.0, 2.12] & $1.4\times10^{-4}$ & 2 & 18.64 min & 0.22 & 409.36 \\
 \textbf{LeakyReLu ($\alpha=0.1$)} & [-7.8, 8.9] & [-0.78, 8.9] & $5.5\times10^{-7}$ & 1 & 5.59 min & 0.12 & 216.72 \\
 \textbf{Squash (8-dim)} & [-1.5, 2.13] & [-1.4, 2.02] & \textbf{$0.0$} & 1 & 5.73 min & 0.12 & 216.72 \\
 \bottomrule
 \end{tabular}
 }
\caption{Different nonlinear operations with varying input parameter sizes replaced by \myName.}
\label{tab:neonnetalgorithmperformance}
\end{center}
\end{table*}

\section{Experimental Methodology}
\label{sec:microarchmethodology}
This section describes the functions transformed by \myName, benchmark workloads, \myName-Net training hyper-parameters, and microarchitecture configurations.

\subsection{Functions transformed by \myName}
We evaluate the accuracy and performance of \myName-Nets for different nonlinear operations in Table~\ref{tab:neonnetalgorithmperformance}. The operations are selected based on our survey in Section~\ref{sec:problem}: softmax (four versions based on different dimensions), square-root, LeakyReLu, and squash. tanh is directly supported by the DLC in the microarchitecture, and sigmoid ($\sigma$) is indirectly supported via the following equation: $\sigma$(z) = (1/2)(tanh(z/2) + 1).
Any operation that is directly or indirectly supported in the microarchitecture is not replaced. 
We restrict our focus to continuous nonlinear operations as neural networks use continuous functions for activation (the function must be differentiable for using back-propagation to train the network~\cite{an1996effects}). 

\subsection{Benchmark Workloads}
\label{sec:benchmark}
We detail the different neural networks in our benchmark in Table~\ref{tab:networkbenchmark}, along with the corresponding unsupported nonlinear operations transformed by \myName. We present the end-to-end system performance results in Section~\ref{sec:systemperfevaluation}.

\subsection{Training Hyper-parameters}
\head{\myName-Net}
\label{sec:neonnethyperparameters}
We use PyTorch~\cite{paszke2017pytorch} as the programming framework. The \myName-Nets are trained on a single GPU (NVIDIA RTX 2070) using the following hyper-parameters: loss function = Mean Square Error (MSE), loss optimization algorithm = Adam~\cite{kingma2014adam}, batch size = 1024, learning rate = $10^{-4}$, weight decay = 0.0001, training epochs = 100, $\epsilon=10^{-4}$. 

\head{Data distributions and pre-processing} \myName-Net training data is collected by executing inference for ten epochs using the \guestName and the original training dataset. The dataset is not normalized before training the \myName-Net to preserve the input value distribution. We use ten epochs for fine-tuning the \guestName after replacing all nonlinear operations with \myName-Nets and report the fine-tuning time in Section~\ref{sec:hostnetworkaccuracyresults}.

\subsection{Microarchitecture configurations}
\label{sec:methodologyRRAMconfigs}
We consider the following microarchitecture configurations for system evaluations:

\head{Digital Logic Circuits (DLC)} The DLC configuration integrates fixed-function digital logic circuits necessary to support different neural networks in the benchmarks. The circuits' area and power consumption values are taken from the respective manuscripts~\cite{chen2020tanh,nilsson2014hardware,Nannarelli2011Radix16CD}. 

\head{Look-Up Tables (LUT)} The LUT configuration integrates lookup tables necessary to support the nonlinear operations in different workloads. We use the resistive crossbars for storing the LUTs.

\head{\myName} The \myName configuration integrates a single digital logic circuit to support the tanh operation. We set the threshold MSE as $10^{-4}$ as it is sufficient for 16-bit precision.%

\head{MLS and OO} Due to fundamental drawbacks associated with MLS and OO (orders of magnitude lower performance and energy efficiency), we do not compare \myName against these methodologies. However, we evaluate \myName's performance against specialized accelerators that use these methodologies to support a particular workload in our benchmark (Section~\ref{sec:contemporaryevaluation}).

 Area and power consumption values for all components are summarized in Table~\ref{tab:microarchparams}. All values have been scaled for the 32 nm process node following the methodology in~\cite{stillmaker2017scaling}. As prior work utilizes 16-bit fixed-point precision, all three configurations (DLC, LUT, and \myName) also work at 16-bit fixed-point precision (input, weight, and output values) for an apples-to-apples comparison. Although resistive crossbars support \texttt{MUL} operation, it requires significant modifications to the wordline drivers~\cite{fujiki2018memory}. Further, multiplying two dynamic values via resistive crossbars requires writing one value to the cells. Write operations incur three orders of magnitude higher energy consumption penalty than read operations. To support multiplications between dynamic values efficiently, we include an optimized multiplier~\cite{mottaghi2009bz} in all configurations. Simulations are performed using a heavily modified version of the NeuroSim toolchain~\cite{chen2018neurosim}.

\begin{table}
  \begin{center}
  \resizebox{1\linewidth}{!}{
  \begin{tabular}{ |c|c|c|c| } 
 \hline
 \textbf{Component} & \textbf{Specification} & \textbf{Power} & \textbf{Area} \\ \hline %
 \multicolumn{4}{|c|}{\textbf{Memory}} \\ \hline
 Subarray & N/A & 24.08 mW & 13120 $um^2$ \\ \hline
 Bank & N/A & 360.79 mW & 484940 $um^2$ \\ \hline
 \hline
 \multicolumn{4}{|c|}{\textbf{Peripheral Circuits}} \\ \hline
 Exponent~\cite{nilsson2014hardware} & \makecell{Power \& Area Optimized} & 7.424 mW & 5017 $um^2$ \\ \hline
 Division/SQRT~\cite{Nannarelli2011Radix16CD} & \makecell{Power \& Area Optimized} & 26.88 mW & 23869 $um^2$ \\ \hline
 Multiplier~\cite{mottaghi2009bz} & \makecell{Power \& Area Optimized} & 4.7 uW & 236 $um^2$ \\ \hline

\end{tabular}
}
\caption{Microarchitectural component power and area values}
\label{tab:microarchparams}
\end{center}
\end{table}

\begin{figure*}[h]
    \centering
    \includegraphics[width = \linewidth]{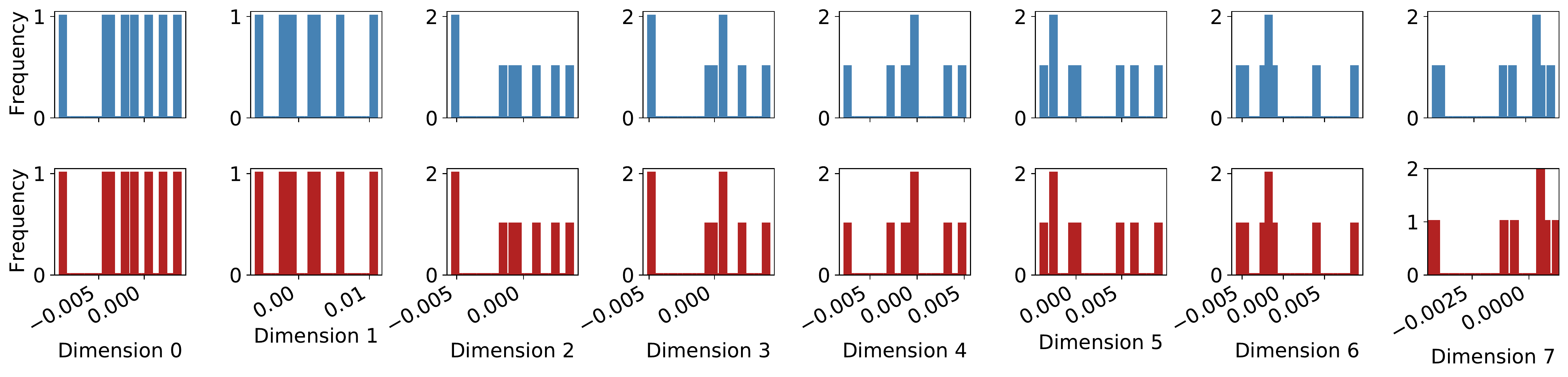}
    \caption{Output value distributions for the squash \myName-Net: ground truth in red and the \myName-Net outputs in blue.}
    \label{fig:softmax_imitation}
\end{figure*}
\section{\myName-Net Evaluations}

This section reports evaluations for \myName-Net accuracy, training time, \guestName's end-to-end accuracy, fine-tuning time, and the trade-off between \myName-Net's size and accuracy.
\subsection{Accuracy}
\label{sec:nearnetaccuracyexperiments}
Table~\ref{tab:neonnetalgorithmperformance} reports the accuracy of the \myName-Nets for different nonlinear operations. To understand the impact of different dimensions, we analyze two different \myName-Nets for softmax: 1000-dimensional from VGG and 1152-dimensional from CapsNet. The 1000-dim softmax \myName-Net requires only one hidden layer and obtains very high accuracy (MSE = $2.62\times10^{-9}$, threshold MSE = $10^{-4}$). In contrast, the 1152-dim softmax requires two hidden layers, resulting in a slight increase in area (13.98\%) and power consumption (14.05\%) of the corresponding \myName-Net compared to the 1000-dim softmax \myName-Net. The SQRT \myName-Net requires the longest training time (18.64m), in contrast to LeakyReLu, with the lowest training time (5.59m). We attribute the differences in training time to the complexity of the respective operations.

\subsection{\myName-Net Structure Generation Time}
\label{sec:nearnettrainingexperiments}
\begin{figure}[h]
    \centering
    \includegraphics[width =\linewidth]{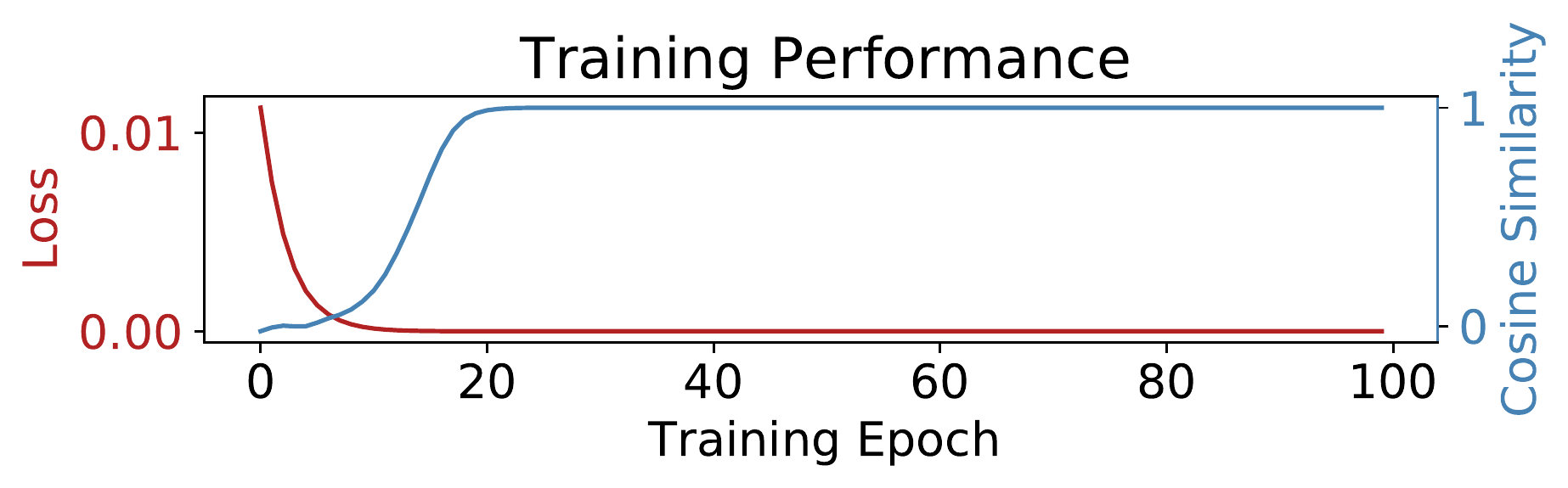}
    \caption{Squash \myName-Net training over 100 epochs completes within 5.731 seconds. Training loss saturates by the 30th epoch (complete in 1.719 seconds).}
    \label{fig:squash_training}
\end{figure}

We report \myName-Net structure generation time for each function in Table~\ref{tab:neonnetalgorithmperformance}. The average time across the benchmark is 12.68 minutes. Figure~\ref{fig:squash_training} displays the training performance curves for an individual \myName-Net (squash). The low training time is attributed to the small size of the \myName-Nets. For example, the squash \myName-Net's structure has three layers with 8, 128, and 8 parameters. We also report the training loss and cosine similarity performance for the training process in Figure~\ref{fig:squash_training}. The performance quickly saturates, and the network demonstrates 1.0 cosine similarity, indicating a very high correlation between the network's output and ground truth vectors.

\subsection{Workload's End-to-End Accuracy}
\label{sec:hostnetworkaccuracyresults}
Table~\ref{tab:networkbenchmark} indicates the end-to-end accuracy loss for different workloads in the benchmark after fine-tuning. The average accuracy loss across the benchmark is -0.54\%, indicating higher accuracy than the baseline. We attribute this observation to the tolerance of neural networks to noise injection~\cite{koppula2019eden}. Fine-tuning the network after replacement allows it to account for the noise. The slight performance improvement is attributed to the regularizing effects of noise injection during training~\cite{noh2017regularizing, an1996effects}.

\head{Fine-tuning time} The average amount of fine-tuning time across the benchmark is 12.88 minutes (Table~\ref{tab:networkbenchmark}). In contrast, training the workloads from scratch requires a few hours on average.

The squash \myName-Net obtains \textbf{0.0 MSE}, implying perfect emulation of the nonlinear operation. To test the impact of 0.0 MSE \myName-Net on the \guestName's (CapsNet) end-to-end accuracy, we replace only the squash operation in the \guestName and skip the fine-tuning step. We observe a 0\% accuracy loss in the \guestName. 
We replace other nonlinear operations in CapsNet (softmax) and observe a 0.99\% end-to-end accuracy loss. However, the \guestName recovers the accuracy loss after fine-tuning and improves upon the baseline accuracy by 1.8\%.

\head{Accuracy across dimensions} We plot the value distributions for each dimension in the squash \myName-Net and the ground truth in Figure~\ref{fig:softmax_imitation} to understand the impact of 0.0 MSE. We observe that the distributions are identical across all eight dimensions. This observation corroborates the 0\% accuracy loss observed in the CapsNet \guestName's end-to-end accuracy when replacing the squash operation with the corresponding (perfect) \myName-Net. 

\head{Fine-tuning Time}
The amount of time needed for fine-tuning each \guestName is as follows: VGG (19 minutes), RNN (6.5 minutes), CapsNet (11 minutes), and transformer (15 minutes). Fine-tuning requires significantly less time than training the \guestName from scratch (requiring multiple hours or days).

\subsection{Exploring \myName-Net's Trade-off Space}
\label{sec:nearnettradeoffs}
We describe the trade-offs associated with \myName.
\subsubsection{Trade-offs between \myName-Net's Size and Accuracy}
\label{sec:epsilonvariation}

We evaluate the performance variation of \myName with an increasing number of neurons per hidden layer. We evaluate the 1000-dimensional softmax obtained from \myName with one hidden layer composed of 128 nodes, with an MSE of 2.62x$10^{-9}$. Increasing the number of neurons in each hidden layer from 128 to 1000 results in a 0.97\% increase in MSE. Further increasing the number of hidden layers from 1 to 3 (each with 1000 nodes) results in a 0.08\% increment in MSE. Although the accuracy improvements are negligible, the \myName-Net's size has increased by $2.5\times$, commensurately increasing the \myName-Net's latency and energy consumption. %

\subsubsection{Discussion against Alternate Machine Learning Algorithms}
Section~\ref{sec:keyinsight} describes the key insights behind \myName: high-performance MAC support in RRAM substrates which helps accelerate neural networks and the capability of neural networks to act as universal function approximators. Using alternate machine learning models such as Support Vector Machines (SVM)~\cite{noble2006svm} and Random Forest Regression (RFR)~\cite{liaw2002randomforest} is not possible as the RRAM substrate cannot execute these models natively. Although RRAM can accelerate linear regression~\cite{ankit2019puma}, such models are not considered viable universal function approximators. Therefore, we believe neural networks are a prime candidate for accelerating nonlinear operations in RRAM.

\subsubsection{Input data distribution shift}
It is possible that the input data distribution shifts compared to the original distribution over which the \myName-Net was trained. The input domain and output range bounds (Section~\ref{sec:inputdomainconstraints}) insulate \myName against this problem to a certain extent. However, this is not a perfect solution, as a significant shift can lead to accuracy loss. In such cases, the \myName-Net must be retrained on the shifted input domain. Note that input data distribution shift is a major problem for neural networks in general, and there is ongoing research in this direction~\cite{shamir2018distribution}.

\section{System Evaluation}
\label{sec:systemperfevaluation}

\subsection{End-to-end evaluations}

We evaluate the end-to-end speedup, power, energy, and resource utilization for the three microarchitecture configurations: DLC, LUT, and \myName.

\begin{figure*}
\begin{subfigure}{.5\textwidth}
  \centering
  \includegraphics[width=.99\linewidth]{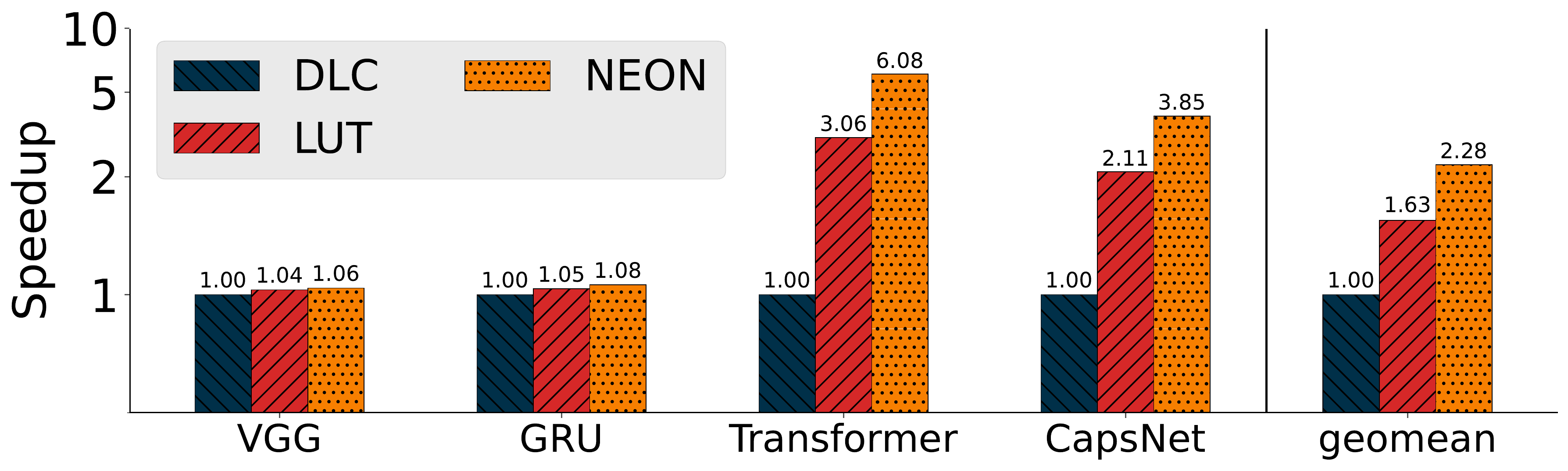}  
   \caption{\small Speedup}
    \label{fig:speedupresult}
\end{subfigure}
\begin{subfigure}{.5\textwidth}
  \centering
  \includegraphics[width=.99\linewidth]{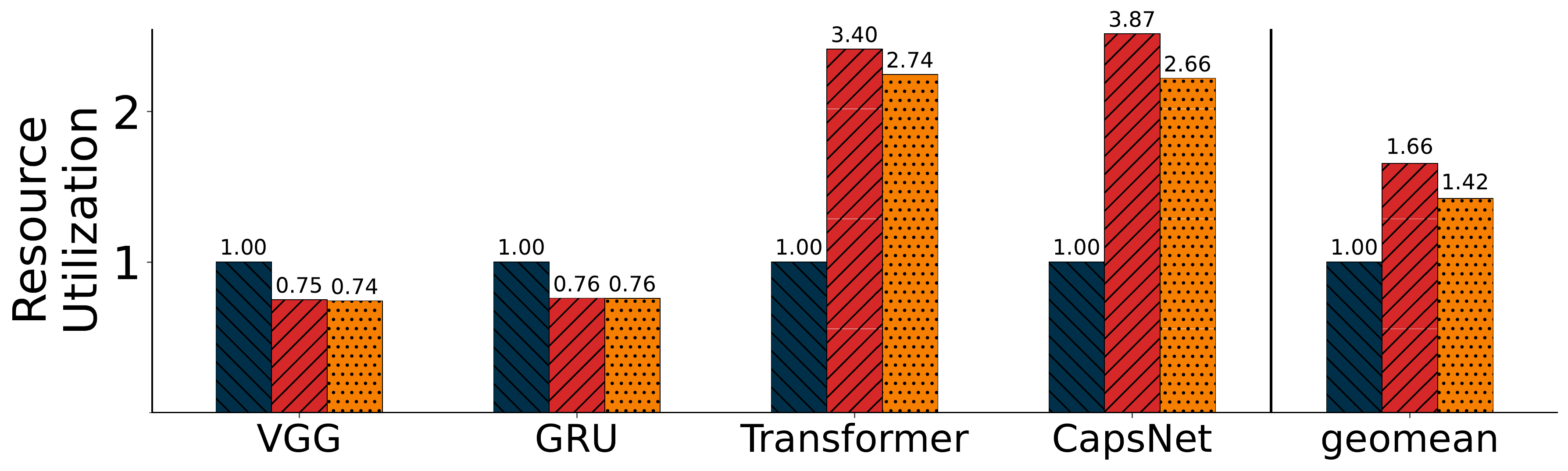}  
  \caption{\small Area utilization}
    \label{fig:arearesult}
\end{subfigure}

\begin{subfigure}{.5\textwidth}
  \centering
  \includegraphics[width=.99\linewidth]{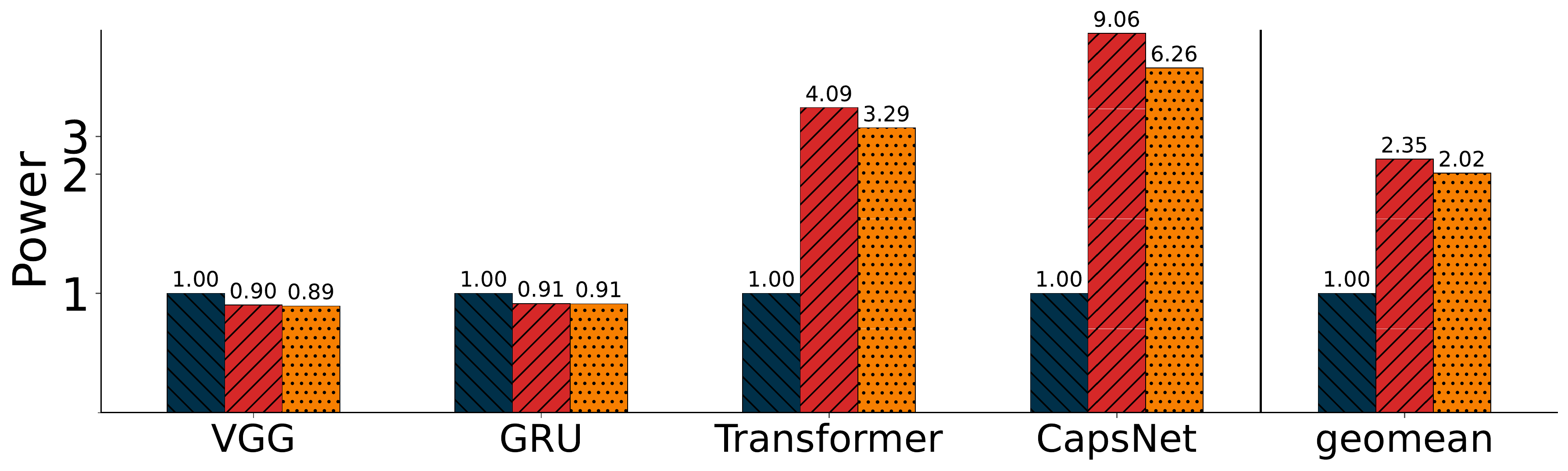}  
  \caption{\small Power Dissipation}
    \label{fig:powerresult}
\end{subfigure}
\begin{subfigure}{.5\textwidth}
  \centering
  \includegraphics[width=.99\linewidth]{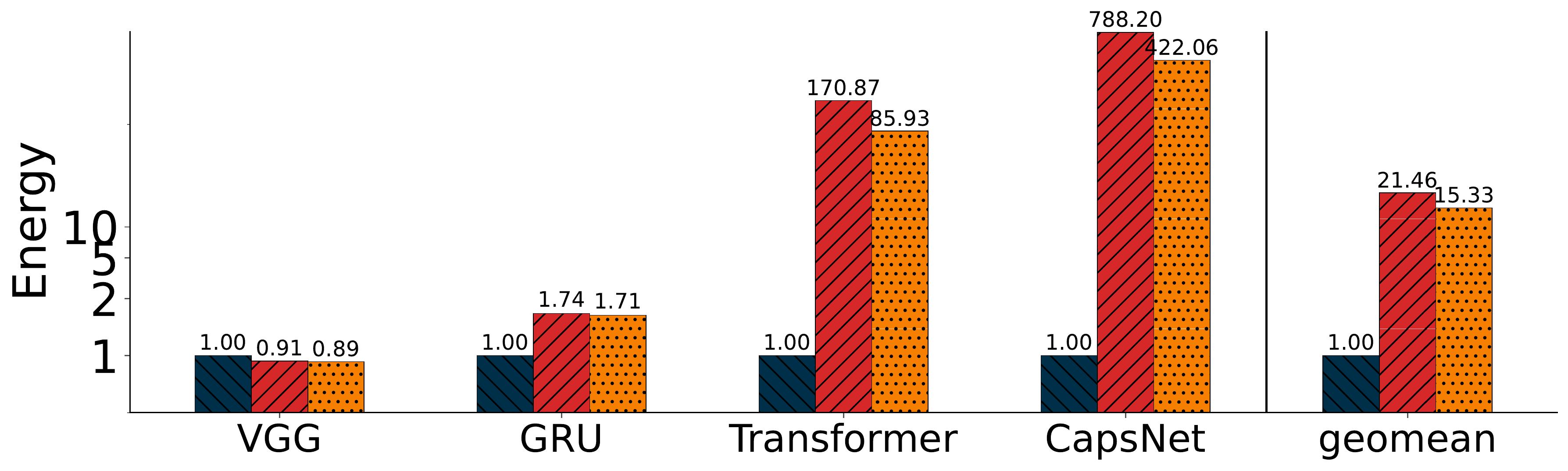}  
  \caption{\small Energy consumption}
    \label{fig:energyresult}
\end{subfigure}
\caption{End-to-end system evaluations across the benchmark neural networks. All values are normalized to the DLC configuration.}
\label{fig:evaluationsoverall}
\end{figure*}

\subsection{Speedup}
\label{sec:speedupresults}
Figure~\ref{fig:speedupresult} shows the speedup normalized to the DLC configuration. \myName consistently provides speedup across the entire benchmark, with a geomean value of 2.28$\times$ and 1.4$\times$ compared to DLC and LUT configurations, respectively. The performance improvement is attributed to the abstraction of long latency operations such as \texttt{EXP} and \texttt{DIV} (in softmax) with MAC and \texttt{tanh} (in \myName-Net). For instance, VGG obtains a modest speedup (1.06x) compared to the transformer (6.08x). The difference is due to a higher fraction of unsupported operations in transformers compared to VGG.

\subsection{Area Utilization}
\label{sec:arearesults}

Figure~\ref{fig:arearesult} compares the area utilization for all three configurations normalized to the DLC configuration. \myName increases the area utilization by $1.42\times$ compared to the DLC configuration. This increase is attributed to the significantly larger resistive crossbars (0.026 mm$^2$) compared to area-optimized digital logic circuits (0.016 mm$^2$ average area). 

\head{Area Utilization for LUTs}
LUTs require 1.17$\times$ more area compared to \myName. The difference is attributed to the value retrieval mechanisms: the neural network stores approximate values in the network's weights (learned via back-propagation). In contrast, LUTs store precise values that require more area.

\subsection{Power Dissipation}
\label{sec:powerresults}
Figure~\ref{fig:powerresult} shows the power dissipation normalized to the DLC configuration. \myName requires 2.02$\times$ higher power compared to the DLC configuration and 1.16$\times$ lower power compared to the LUT configuration. The difference with respect to DLC is attributed to the difference in power dissipation of fixed-function circuits (10.28 mW on average) compared to resistive crossbars (24.08 mW on average). The power dissipation of resistive crossbars is dominated by the ADCs (16 mW). Lowering ADC power can help improve \myName's power efficiency~\cite{murmann2015race}.

\subsection{Energy Consumption}
\label{sec:energyresults}
Figure~\ref{fig:energyresult} shows the energy consumption normalized to the DLC configuration. Despite a reduction in the execution time, higher area utilization and power dissipation lead to higher energy consumption (15.33$\times$ higher than DLC and 1.4$\times$ lower than LUTs). Workloads with a higher fraction of transformed operations (Transformer and CapsNet) report significantly higher energy consumption.

\subsection{Operator Scaling}
\label{sec:inputoperatorscalingresult}
\begin{figure}[h]
  \centering
  \includegraphics[width=\linewidth]{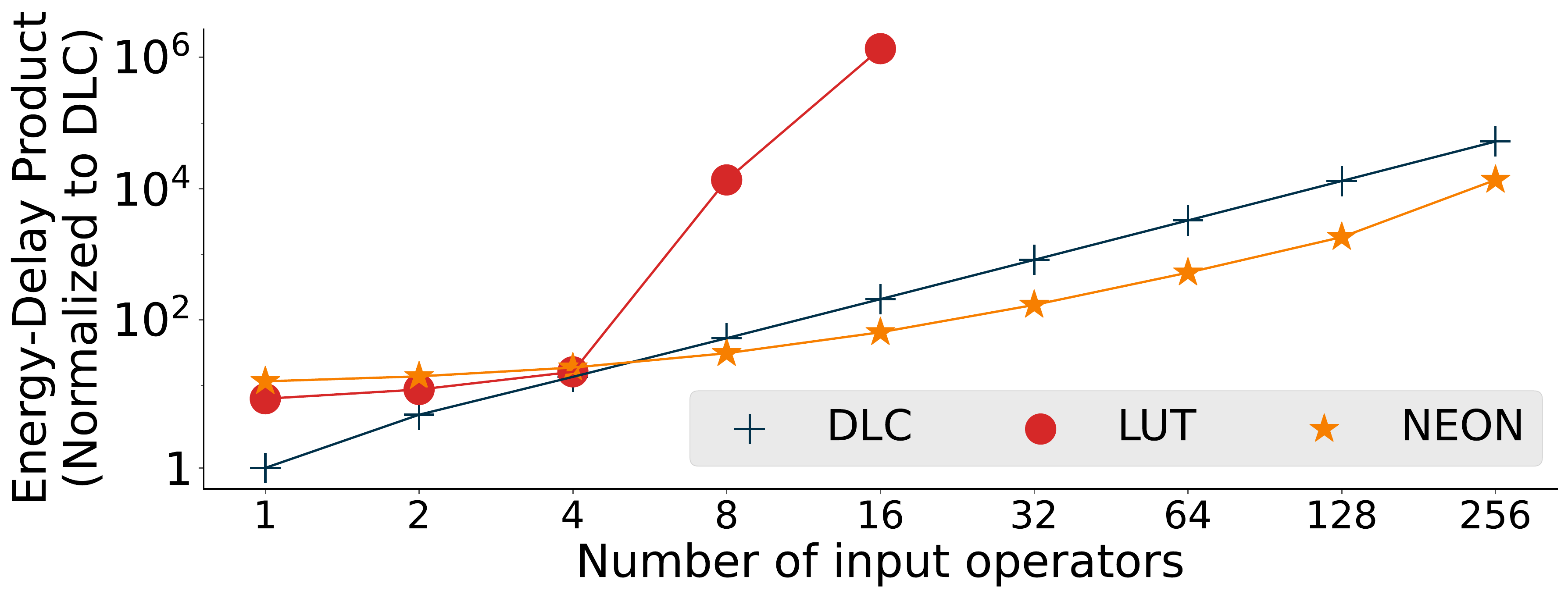}
  \caption{EDP Scaling with the number of input operators. LUTs are unable to scale without untenable EDP overheads.}
    \label{fig:inputoperatorscalingresult}
\end{figure}

Figure~\ref{fig:inputoperatorscalingresult} shows the energy-delay product (EDP) for all configurations normalized to DLC as we scale the number of input operators. We observe that digital logic offers lower EDP for a single input operator compared to \myName-Net. This observation is attributed to the higher cost of an entire subarray dedicated to execute the \myName-Net. However, as we increase the number of input operators, we observe that the EDP for digital logic scales linearly due to the fixed cost increments needed to support each new input value. In contrast, \myName-Net yields sub-linear EDP scaling by using more rows in the dedicated subarray. It is worth noting that \myName-Net shows an increase in the slope of the curve from 128 to 256 inputs. This observation is attributed to the addition of an additional subarray upon exceeding the capacity of the first one.

\subsection{\myName-Net Initialization Energy Consumption}

\myName-Net subarrays must be initialized via additional memory writes before deploying the system. We consider the initialization cost of \myName-Nets and compare it to a single inference call's energy requirement for the corresponding workload. \myName-Net initialization consumes on average 3.54\% of a single inference call's energy consumption.

\subsection{Comparison against Relevant Prior Work}
\label{sec:contemporaryevaluation}
ReTransformer~\cite{yang2020retransformer} proposes a design for accelerating transformers in RRAM. \myName achieves $29.56\%$ speedup over ReTransformer. Long et al.\cite{long2018reram} propose a design for accelerating RNNs in RRAM using flexible logic circuits. \myName achieves $11.51\times$ speedup and $14.58\times$ energy reduction over \cite{long2018reram}. Zhang et al.~\cite{Zhang2021AUR} propose a CORDIC-based microarchitecture supporting different nonlinear operations in RRAM. \myName obtains $87.09\times$ higher performance over \cite{Zhang2021AUR}. 

\section{Related Work}

This section discusses prior efforts to support nonlinear operations in RRAM and neural network-based code approximation.

\subsection{RRAM for Accelerating Neural Networks}
Few prior works propose RRAM substrates to support different neural networks and machine learning workloads. 
Ankit et al.\cite{ankit2019puma} propose PUMA, which relies on lookup tables for supporting non-native operations. We demonstrate that LUTs are area-inefficient compared to \myName.
Other works~\cite{Ji2017BridgeTG, ji2020reduced} look at designing RRAM-based neural network accelerators in the context of spiking neural networks (SNNs). However, SNNs require different hardware support structures (spike generator and accumulator in contrast to DACs and ADCs). Consequently, these works require significant hardware customization (e.g., FPGA-like interconnects~\cite{ji2019fpsa}). \myName focuses on supporting different nonlinear operations in ADC-based RRAM accelerators.
Zhang et al.~\cite{Zhang2021AUR} propose an RRAM substrate that supports different operations by relying on the CORDIC~\cite{volder1959cordic} algorithm for transcendental functions. We demonstrate that \myName is significantly faster than this proposal (Section~\ref{sec:contemporaryevaluation}). Further, PUMA~\cite{ankit2019puma} corroborates our hypothesis that a sufficiently accurate CORDIC unit is infeasible in practice due to a large area requirement and high implementation complexity.
Huang et al.~\cite{3dmemristor} propose a 3D-RRAM microarchitecture for accelerating capsule networks. However, their proposal offloads all nonlinear operations to the host. In contrast, \myName supports the nonlinear operations \emph{natively} in RRAM.

\subsection{Neural network-based code approximation}
\label{sec:improvementsoverpriorwork}
Esmaeilzadeh et al.\cite{esmaeilzadeh2012neural} replace manually identified code sections in general-purpose workloads with a human-trained neural network. In contrast, \myName automatically replaces unsupported nonlinear operations in neural network workloads to improve amenability on the target Processing-in-Memory substrate (RRAM). \myName executes the replacement of non-linear functions into neural networks automatically using a reproducible process. We detail the differences as follows:

\head{Automatically identifying code segments for transformation}
Prior work~\cite{Moreau2015SNNAPAC, Yazdanbakhsh2017AxBenchAM, yazdanbakhsh2018dram} relies on the programmer to manually identify and annotate suitable code regions for replacement. \myName overcomes this limitation by leveraging information available at compile-time from the \guestName's execution graph. 

\head{Automated neural network definition and training}
Prior works~\cite{Yazdanbakhsh2015NeuralAF, Yazdanbakhsh2015AxilogLS} rely on the programmer's expertise in machine learning to design the replacement neural network structure and train it for high accuracy. \myName overcomes these limitations by automating the network structure definition and training process.

\section{Extensions and Future Work}
This section discusses the extensibility of \myName to different non-volatile memory-based PIM substrates, future directions for \myName, and manufacturing challenges for integrating DLC in memory microarchitectures.
\subsection{Applicability of \myName to Different Non-Volatile Memory-based Processing-in-Memory Substrates}
\label{sec:applicabilityexpansion}

Emerging Non-Volatile Memory (NVM) technologies such as Phase Change Memory (PCM)~\cite{Wong2010PhaseCM, le2020overview, burr2016recent,lee2009architecting, Qureshi2009ScalableHP, Joshi2020AccurateDN,Burr2015ExperimentalDA, xu2014architecting} and Spin-Transfer Torque Magnetic RAM (STT-MRAM)~\cite{Shi2020PerformancePO, 8119280} have gained attention as novel PIM substrates. These substrates perform operations on data values stored in the memory subarrays, using bitline-based computation mechanisms~\cite{8490883, tsai2018recent, li2016pinatubo} often organized similar to RRAM microarchitectures. %
Orthogonal to the underlying NVM technology choice, \myName provides a novel approach to support nonlinear operations in different substrates designed for accelerating neural network workloads.
Although \myName is presented and evaluated in the context of RRAM in this paper, we believe that it is easily extensible to other substrates. Evaluating the feasibility and performance of \myName for different substrates is left for future work.

\subsection{Future Directions for \myName}
\head{Neural Architecture Search (NAS)}
\label{bg:neuralarchitecturesearch}
\myName enables generalizable support for nonlinear operations in RRAM while opening a new research problem: how to find high-accuracy and high-performance neural networks for emulating different nonlinear operations? NAS~\cite{zoph2016neural, zoph2018learning, Baker2017DesigningNN, Real2017LargeScaleEO, Xie2017GeneticC,li2020neural, jiang2019accuracy} offers one potential research direction towards this goal. NAS transforms the network design process into a search space exploration using a gradient method (e.g., back-propagation or reinforcement learning)~\cite{jin2019auto,gong2019autogan}. A loss function guides the search based on accuracy and performance metrics. NAS optimizes for two orthogonal problems in parallel – designing the network's structure (including the size, number, and type of layer) and optimizing its trainable parameters (weights)~\cite{cai2018proxylessnas, ying2019bench, nayman2019xnas,liu2018progressive}. NAS-generated neural networks often significantly outperform manually designed networks in accuracy and performance~\cite{tan2019mnasnet, wu2019fbnet,wan2020fbnetv2}. However, considering the complexity of realizing NAS in practice, this direction is left for future work.

\subsection{Manufacturing Challenges for Integrating Digital Logic in Memory Microarchitectures}
\label{sec:manufacturingchallenges}

The manufacturing processes for integrating a large amount of digital logic on RRAM substrates remain an open challenge~\cite{Wu2017DeviceAC,wong2015memory,turkyilmaz2012rram, tang2020rram, tanachutiwat2010fpga, 9097632}. Memory microarchitectures are optimized for density in contrast with performance-optimized logic process~\cite{Zahoor2020ResistiveRA, Levisse2018RRAMCA, Lee2016ANR}. Consequently, integrating general-purpose cores or FPGA units in memory substrates presents significant challenges. Further, programming such systems requires complex instructions that are generally not a part of memory ISAs~\cite{Armstrong2019ISASF}.

\section{Conclusion}
We propose \emph{\myName}, a novel hardware/software co-design methodology to efficiently support different nonlinear operations in RRAM. \myName enables RRAM to overcome the fundamental restrictions on supported operations by exploiting key strengths of the substrate. Further, it improves the end-to-end system performance compared to the DLC and LUT methodologies across different neural networks. 
We hope this work opens a new research direction in RRAM microarchitecture design to enable support for different operations without additional computation structures.

\bibliographystyle{unsrt} 
\bibliography{refs}

\end{document}